\documentclass[10pt]{wlscirep}
\usepackage[utf8]{inputenc}
\usepackage{float}
\usepackage[T1]{fontenc}
\usepackage[draft]{changes}
\title{Quantifying spatial homogeneity of urban road networks via graph neural networks\footnote{\small This study was in principle accepted by Nature Machine Intelligence on Nov. 26, 2021. We are now performing the format revision.}} %.. good. 2

\author[1]{Jiawei Xue}
\author[2]{Nan Jiang}
\author[3]{Senwei Liang}
\author[3]{Qiyuan Pang}
\author[1]{Takahiro Yabe}
\author[1,*]{Satish V. Ukkusuri}
\author[2,4,*]{Jianzhu Ma}

\affil[1]{Lyles School of Civil Engineering, Purdue University, West Lafayette, Indiana, USA}
\affil[2]{Department of Computer Science, Purdue University, West Lafayette, Indiana, USA}
\affil[3]{Department of Mathematics, Purdue University, West Lafayette, Indiana, USA}
\affil[4]{Institute for Artificial Intelligence, Peking University, Beijing, China }

\affil[*]{Correspondence should be addressed to: sukkusur@purdue.edu, majianzhu@pku.edu.cn}

\begin{abstract}
Quantifying the topological similarities of different parts of urban road networks (URNs) enables us to understand the urban growth patterns. While conventional statistics provide useful information about characteristics of either a single node's direct neighbors or the entire network, such metrics fail to measure the similarities of subnetworks considering local indirect neighborhood relationships. In this study, we propose a graph-based machine-learning method to quantify the spatial homogeneity of subnetworks. We apply the method to 11,790 urban road networks across 30 cities worldwide to measure the spatial homogeneity of road networks within each city and across different cities. We find that intra-city spatial homogeneity is highly associated with socioeconomic statuses such as GDP and population growth. Moreover, inter-city spatial homogeneity obtained by transferring the model across different cities, reveals the inter-city similarity of urban network structures originating in Europe, passed on to cities in the US and Asia. Socioeconomic development and inter-city similarity revealed using our method can be leveraged to understand and transfer insights across cities. It also enables us to address urban policy challenges including network planning in rapidly urbanizing areas and combating regional inequality.
\end{abstract}
\begin{document}
\maketitle
\thispagestyle{empty}
%\noindent Keywords: network homogeneity, urban studies, road networks, urban development, machine learning
\section*{Introduction}
%1. There are many measurements of urban road networks.
%2. The current measurements do not consider the homogeneity of urban road networks. 
%3. Homogeneity of road networks has some connections with socioeconomic factors: consistency of urban planning. 

%4. The homogeneity reveals high-resolution urban process. First, infrastructure-society. which exceeds the traditional aggregated analysis of infrastructure and social activities. Second, inter-city similarity. 

%5. What do we.
%6. The concise contribution of this study: fully mine the road network structure information using graph neural networks, and reveal strong correlation between infrastructure and social-economic status.

Road networks have sprawled due to rapid urbanization over the past decades, especially in developing countries. Urban road networks (URNs) play a central role in efficient social communication \cite{sun2013understanding,roth2011structure}, the movement of goods\cite{steadieseifi2014multimodal}, and are complicated due to their interdependence with the population \cite{bettencourt2007growth, arcaute2015constructing}, human mobility \cite{xu2020deconstructing,snellen2002urban}.
There are underlying signatures that reveal the interplay between the structure and functionality (such as road usage\cite{wang2012understanding}, network dynamics \cite{zhan2017dynamics,li2015percolation}, traffic congestion \cite{saberi2020simple,ccolak2016understanding,zhang2019scale}, and land use \cite{foley2005global}). Urban scientists have exploited various topological features, such as road length distribution \cite{strano2017scaling}, relative angle \cite{molinero2017angular}, degree distribution \cite{kalapala2006scale,porta2006network}, betweenness centrality \cite{crucitti2006centrality,kirkley2018betweenness}, clustering coefficient \cite{jiang2004topological}, block shape \cite{louf2014typology}, and travel routes morphology \cite{lee2017morphology}. At an aggregate level, "scaling laws" are widely investigated theories that either characterize the distribution of road network topological features\cite{strano2017scaling,kalapala2006scale,zhan2017dynamics} or present the scale-invariant relationships between road networks and other urban indicators\cite{masucci2015problem,lammer2006scaling,bettencourt2007growth,depersin2018global}. Many of these metrics and laws either focus on the direct neighbor node-level characteristics (e.g., node degree, clustering coefficient) or the aggregate-level characteristics of the entire network (e.g., betweenness, scaling laws) and are thus insufficient to capture indirect neighborhood node (such as two-hop neighbors) relationships. Here, node $u$ is an indirect neighborhood of node $v$ if and only if the shortest path between $u$ and $v$ has a length larger than 1.

Local network metrics such as local betweenness\cite{thadakamalla2005search,jeong2007low,ahmadzai2019assessment,nigam2021local} and local closeness\cite{porta2012street,ahmadzai2019assessment,mahyar2019compressive} inherit the same formulas from the global metric and focus on the subnetwork surrounding a specific node. Subgraph theories such as motifs \cite{schneider2013unravelling,dey2019network} and higher-order interactions\cite{benson2018simplicial} mine the structure and interaction patterns among a group of nearby nodes. Both of them are capable of describing the indirect neighborhood node relationships. However, they do not directly quantify the similarity of subnetworks due to the lack of subsequent distance (or similarity) functions. Furthermore, many distance functions (e.g., Euclidean distance) are manually designed and return unbounded values from 0 to $\infty$, which contradicts the fact that existing similarity measures in statistics (e.g., Pearson correlation) are mostly bounded (Please find discussions in Supplementary Section 3.1). Hence, we currently lack a network metric that measures the similarities of local topological patterns across subnetworks considering indirect neighborhood node relationships (\textbf{Fig. \ref{fig:1}$a$}).
In this study, we use \textit{spatial homogeneity} to describe the similarities between subnetworks capturing the indirect neighborhood relationships.
%For urban road networks, quantitatively understanding the homogeneity of URNs by measuring the similarity of different parts of the network (subnetworks) enables us to understand the growth and generative processes of these networks.

The spatial homogeneity of a URN represents the level of similarity in the topological signatures of intersection connections across different regions within a city, which may have been constructed in distinct periods with varying historical contexts. Intersections and neighborhood blocks constructed under unified road planning standards would have high spatial homogeneity, while roads planned under varying infrastructure policies would have low spatial homogeneity. Recall that the spatial homogeneity is defined from the multi-hop neighborhood node relationships, which embody local road network features that moving agents would perceive during daily travels. Hence, the spatial homogeneity is likely to reveal urban knowledge by extracting multi-hop neighborhood node information from URNs. On the one hand, the spatial homogeneity of URNs is promising to connect with socioeconomic development thanks to the relationships between urban planning, economic activities, and spatial homogeneity. These insights display nuanced interactions between infrastructure networks and socioeconomic environment \cite{chandra2000does, molinero2021geometry}. On the other hand, if the scope of spatial homogeneity is extended from subnetworks within one city to subnetworks across multiple cities, the spatial homogeneity captures inter-city similarity \cite{currid2010two} of the road network structure and socioeconomic processes. Such observations are rather valuable to analyze the interactions of URNs from cities in developed and developing countries in a global manner. 

To derive the aforementioned urban insights from URNs, the foremost task is to define the spatial homogeneity capturing indirect neighborhood node relationships that are missing in existing URN metrics. %Addressing these limitations necessitates a novel approach to capture the multi-hop relations in road networks. 
We overcome this challenge using graph neural networks (GNNs), which is an innovative model and a variant of deep neural networks designed for graph-structured data. GNNs have already achieved tremendous success in a diverse array of graph-based employments, such as drug discovery \cite{cheng2019network}, social friendship prediction \cite{jalili2017link}, social network embedding \cite{lerique2020joint}, human mobility computation \cite{ren2014predicting}, and visual question answering \cite{teney2017graph}. The multi-hop message passing mechanism in GNNs effectively captures multi-hop node relationships \cite{wu2020learning} in URNs. We claim there is a fundamental connection between the spatial homogeneity of URNs and the predictability of road links using GNN models, as they both measure the similarity of topology between subnetworks and the whole network. In addition, neural newtorks (NNs) have been successfully leveraged to not only predict urban factors\cite{gebru2017using} but also reveal urban knowledge \cite{abitbol2020interpretable,kempinska2019modelling}. Consequently, we train GNN models to predict the existence of road links on URNs from 30 cities in the US, Europe, and Asia, and then use the prediction accuracy metrics to define the road network spatial homogeneity within and across cities. Despite using only the network information, the road network spatial homogeneity is found to have strong correlations with various socioeconomic factors, allowing us to further interpret the prediction results. 

Remarkably, we find that road network spatial homogeneity uncovers profound socioeconomic development and inter-city urban similarity patterns that are invisible to existing road network metrics (\textbf{Fig. \ref{fig:1}$a$}). Note that we use two standard indexes, \textit{GDP} and \textit{population growth} to characterize the socioeconomic development. We compute the intra-city spatial homogeneity and find that cities with high GDP and low population growth have significantly higher spatial homogeneity than other cities. To quantify inter-city spatial homogeneity, we establish the City Homogeneity Transfer Matrix (CHTM) by measuring the transferability of the GNN models among different cities. Results from CHTM suggest that city clusters are highly consistent across different areas (the US, Europe, and Asia) and reveal far-reaching global urban dissemination patterns \cite{peng2011urbanization}. 

Concisely speaking, the main contribution of this study is to mine fine-grained local urban road network information using GNNs and present its significant connections with intra-city development and inter-city urban similarity. Our goal is to understand entangled relationships between infrastructure and urban socioeconomic processes. Our metrics and findings are particularly valuable to various academic disciplines and stakeholders in practice (\textbf{Fig. \ref{fig:1}$a$}): 1) intra-city spatial homogeneity helps urban regional and social scientists assess the spatial infrastructure equity \cite{hanson2004geography}, 2) inter-city spatial homogeneity serves as a quantitative measure of road network similarity to support various cross-city transfers of policies \cite{cook2008mobilising} such as autonomous vehicle policy, and accident prevention policy, especially from cities in developed countries to developing countries, and 3) inter-city urban insights document the urban historical patterns which are useful for urban archaeologists. 

\section*{Results}

\subsection*{Measuring spatial homogeneity of 30 major global cities using GNN models}
URNs are modeled as graphs, where intersections and road segments are denoted by nodes and links, respectively (\textbf{Fig. \ref{fig:1}$b$}). We quantify the spatial homogeneity of the network by taking the $F1$ score (prediction accuracy) of the “link prediction” problem \cite{ghasemian2020stacking,clauset2008hierarchical,stanfield2017drug}, where we predict the presence of a link based on the structural roles of its two endpoints, using a graph neural network model (R-GCN model)\cite{schlichtkrull2018modeling}, as shown in \textbf{Figs. \ref{fig:1}$cd$} (\textbf{Methods}). The intuition of selecting this task is that a URN with high spatial homogeneity is likely to contain many links that can be recovered (predicted) by the rest of the graph. We perform link prediction on 11,790 URNs from 30 cities in the US, Europe, and Asia (Supplementary Table 1) and achieve the overall average $F1$ score of 0.42 (\textbf{Fig. \ref{fig:1}$e$}). 

We first show that the spatial homogeneity metric is a composite measure that captures various existing network statistics, including the average degree and betweenness. We find that cities with low average betweenness and high average degrees (e.g., New York, Chicago, and San Antonio) have higher spatial homogeneity ($P$-values$<$0.001 under one-sided $t$-tests) when the 30 cities are evenly distributed into two groups based on their average betweenness and average degree values (Supplementary Fig. 17). This is due to intersections with more road segments receiving more neighborhood information via the message passing mechanism in the GNN model, resulting in higher predictability. In contrast, a road segment with high betweenness (such as a bridge) attracts many shortest-path flows and is distinct from their neighboring road segments, and therefore provides low predictability.

Next, we interpret the prediction results by connecting them with road network clustering from existing works. As shown in \textbf{Fig. \ref{fig:1}$e$}, although the overall performance is much higher than the random, the prediction results show a certain variance among different cities. A higher $F1$ score implies that the subnetworks are more predictable using the global information, but this understanding is insufficient to connect with the existing urban science theories. Hence, we investigate whether different topology types of URNs, defined by existing urban science studies\cite{barrington2019global,barrington2020global}, have significantly different $F1$ scores. In this way, we directly connect our new metric to the well-defined URN types, which are constructed based on multidimensional representations of existing network factors. Note that we perform the clustering on URNs based on multidimensional representations of existing network metrics rather than $F1$ scores, because the former provides more complete topology information than the $F1$ score.
For each one of 11,790 URNs, we calculate 11 existing network metrics and encode each URN as an 11-dimension vector point whose elements are these metric values (\textbf{Fig. \ref{fig:2}$a$}, \textbf{Methods}). Next, on the 11-dimension real number space, we cluster these 11,790 points into 4 types using the $K$-means algorithm.
Results show that Type 1 URNs have the largest average degree (3.11, Types 2 to 4: 2.78, 2.64, 2.60) and the largest fraction of “Degree$\geq$4” nodes (37.8\%, Types 2 to 4: 19.6\%, 20.6\%, 17.0\%). Type 2 URNs are characterized with a large proportion of “Degree=3” nodes (47.7\%, Types 1, 3, 4: 41.2\%, 41.5\%, 40.6\%). The key features defining Type 3 URNs are the “Degree=1” nodes and dead-end edges. Type 4 URNs are distinct, with the largest values of circuity and fraction of bridges. Based on these dominant features, we label Types 1 to 4 as “\textit{Grid}”, “\textit{Degree-3}”, “\textit{Irregular grid}”, and “\textit{Circuitous}”, respectively. For each type, we visualize 12 representative URNs in the US, Europe, and Asia in Supplementary Figs. 20-23. The differences in $F1$ scores across cities (\textbf{Fig. \ref{fig:2}$b$}) can be explained using the road types. \textbf{Fig. \ref{fig:2}$c$} shows the significant performance disparities across these four types of URNs. $F1$ scores associated with “Grid” type (average: 0.526) are significantly higher than “Degree-3”, “Irregular grid”, and “Circuitous” (average: 0.413, 0.341, and 0.314) with $P$-values$<$10$^{-37}$ under one-sided $t$-tests, suggesting more predictable patterns existing in the “Grid” type. We thus hypothesize that a city with a large proportion of “Grid” type could have high predictability. To test the hypothesis, we calculate the URN type composition for each city (\textbf{Fig. \ref{fig:2}$d$}; Supplementary Fig. 19). As expected, the average $F1$ score of a city is positively correlated with the proportions of the “Grid” type (Pearson correlation=0.702). To evaluate the robustness of the clustering results, we randomly select 8, 9, and 10 features among the 11 metrics and re-run the clustering results. We find that the URN clustering results are highly consistent. For instance, at least 93.1\%, 75.4\%, 80.2\%, and 84.4\% of URNs from Types 1 to 4 have the same memberships when 3 features are removed (Supplementary Fig. 15).

To further dissect the spatial homogeneity metric, we perform the principal component analysis (PCA) on 11-dimension vector representations of URNs. The first principal component PC1 encodes a small average degree, large circuity and dendricity, explaining 48\% of data variability (Supplementary Table 4). PC1 was defined as Street-Network Disconnectedness index (SNDi) and investigated in the previous studies \cite{barrington2019global,barrington2020global}. For brevity, we refer to PC1 as Network Irregularity (NI). NI is shown to be positively related with circuity and dendricity (the fraction of bridge edges and dead-end edges) and negatively related to the average degree. \textbf{Fig. \ref{fig:2}$e$} visualizes the NI and $F1$ scores for testing URNs in 30 cities and presents the stunning correlation: $F1$ score is negatively correlated to NI (Pearson correlation=$-0.484$), implying that a URN with a larger NI is less predictable. This negative relationship is more prominent in the US (Pearson correlation=$-0.563$) than in Europe, Asia (Pearson correlation=$-0.269, -0.430$, Supplementary Fig. 24). The US cities have a relatively short history of large-scale population agglomeration, and their road networks were built under standardized land ordinance policies, which trace back to the grid plan in Philadelphia, PA in the early 18th century \cite{hammack1984philadelphia}. In contrast, megacities in Europe and Asia (e.g., London, Beijing) experienced long periods of both top-down planning policies from authority and self-organization from residents, which simultaneously carved the urban prototype \cite{barthelemy2013self}. As illustrated in \textbf{Figs. \ref{fig:2}$f$-$h$}, Chicago has a prevalent distribution of grid-like blocks, while Paris and Singapore are occupied by roads with irregular blocks. 

\begin{figure}[H]
\centering
\begin{minipage}{1.00\textwidth}
    \centering
  \includegraphics[width=16cm]{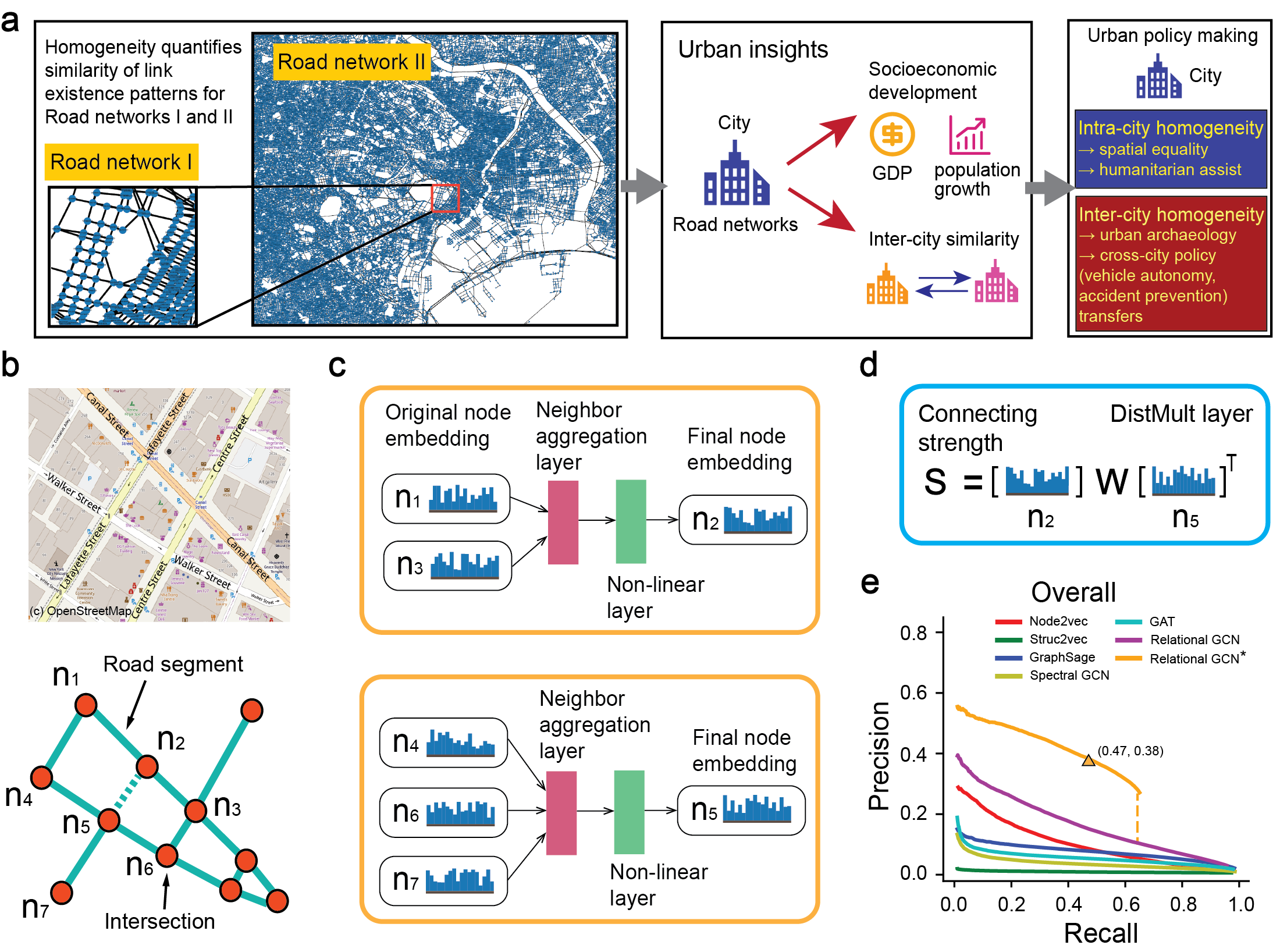}
  \label{fig:1}
\end{minipage}
\caption{\footnotesize
\textbf{Spatial homogeneity.} 
($\textbf{a}$) Description of the spatial homogeneity. Spatial homogeneity measures whether the part network shares analogous link existence patterns with the entire network.  ($\textbf{b}$) A road network in Manhattan, New York. We model intersections as nodes, road segments as links in a graph neural network (GNN). ($\textbf{c}$) Message-passing between two layers in the GNN. The node embeddings (representations) of nodes 2 and 5 are non-linear aggregations of their neighborhood nodes’ (nodes 1, 3, nodes 4, 6, 7) embeddings in the lower layer (see \textbf{Methods} for details). ($\textbf{d}$) Connecting strength of a pair of nodes. It is calculated as a weighted dot product of the representations of two endpoint nodes ($W$ is a diagonal matrix). If the sigmoid connecting strength is above a threshold $\delta$, then we predict the link between nodes 2 and 5 exist. ($\textbf{e}$) The model performance represented by the precision-recall curves of the link prediction for the road network samples in 30 cities (Supplementary Table 1) under different GNN models. We define the road network spatial homogeneity as the $F1$ score of the best GNN model with a well-tuned $\delta$. The triangle represents the model and $\delta$ with the best performance, which achieves the average recall and precision of 0.47 and 0.38, then the average $F1$ score of 0.42. We later see significant discrepancies among 30 cities in terms of the $F1$ score (see \textbf{Results} for details) and report a wide variety of urban insights based on these $F1$ scores.}
\label{fig:1}
\end{figure}

\begin{figure}[H]
\centering
\begin{minipage}{1.00\textwidth}
    \centering
  \includegraphics[width=16cm]{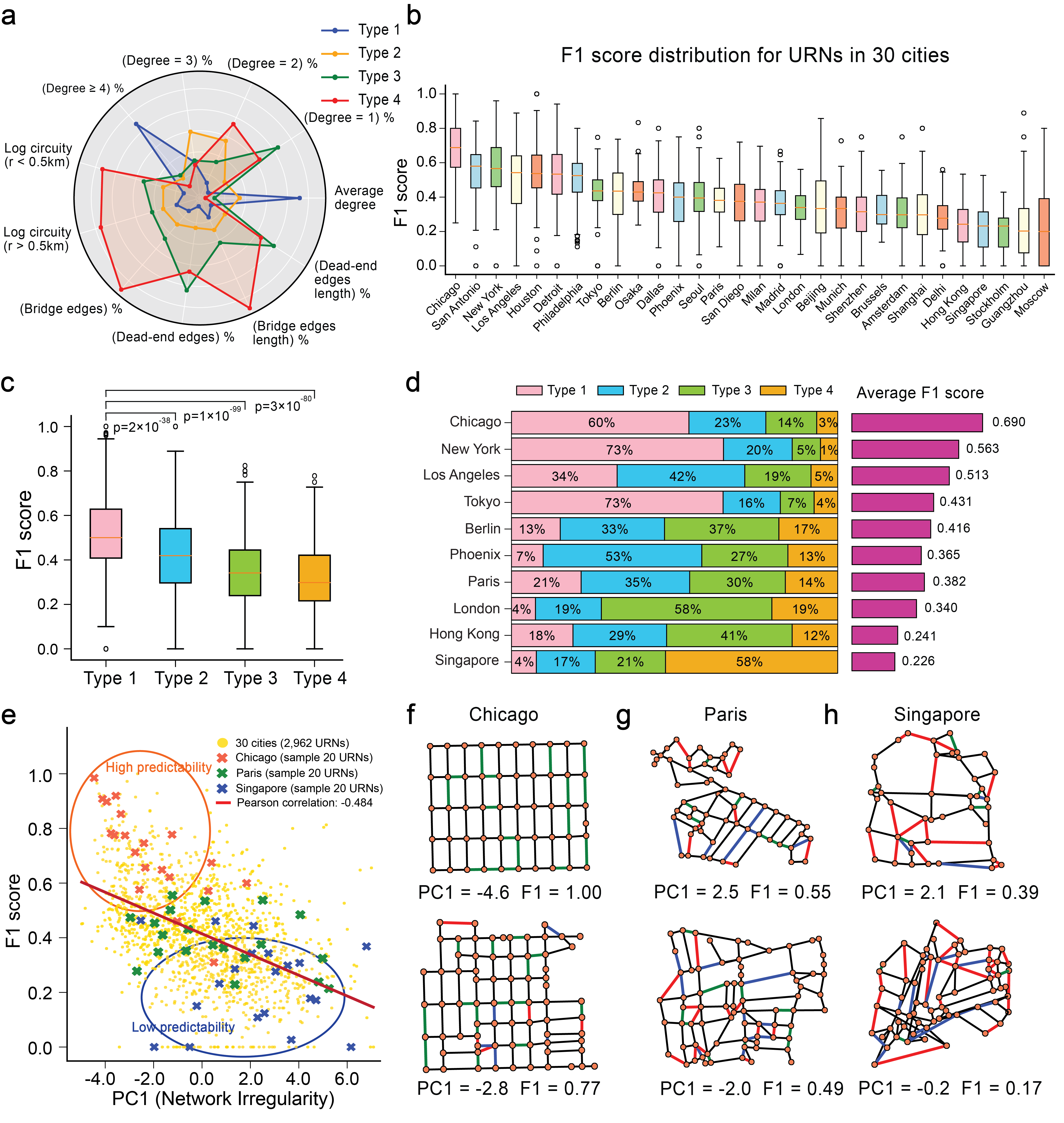}
  \label{fig:2}
\end{minipage}
\caption{\footnotesize
\textbf{Interpretation of the spatial homogeneity.} ($\textbf{a}$) Four urban road network types. This figure shows centroid coordinates of 4 types of URNs. For each road network, we calculate 11 different network metrics (\textbf{Methods}), perform the clustering on these 11-dimension vectors, and finally get the 4 types of URNs. ($\textbf{b}$) $F1$ score distribution. For each city, we train a GNN model and apply the model to many testing road network samples in the same city, then we get a distribution of $F1$ scores for all testing samples. We find Chicago achieves the highest average $F1$ score. ($\textbf{c}$) Comparison of $F1$ score across 4 types of URNs. We find Type 1 URN achieves a significantly higher $F1$ score than the other 3 types. ($\textbf{d}$) Type compositions and $F1$ scores of URNs in 10 cities. There is a general tendency that the average $F1$ score increases with the ratio of Type 1 URNs in the city (see the statistical evidence in \textbf{Results}, 30 cities' results in Supplementary Fig. 19). ($\textbf{e}$) Relationship between the $F1$ score and the PC1 values. $F1$ score decreases when the PC1 value increases, suggesting that regular URNs are more likely to be homogeneous. ($\textbf{f}$-$\textbf{h}$) Samples of link prediction results. Green links denote true positive (links exist in both reality and prediction), red links denote false positive (links exist in prediction but not in reality), and blue links denote false negative (links exist in reality but not in prediction).
}
\label{fig:2}
\end{figure}

\begin{figure}[H]
\centering
\begin{minipage}{1.00\textwidth}
    \centering
  \includegraphics[width=16cm]{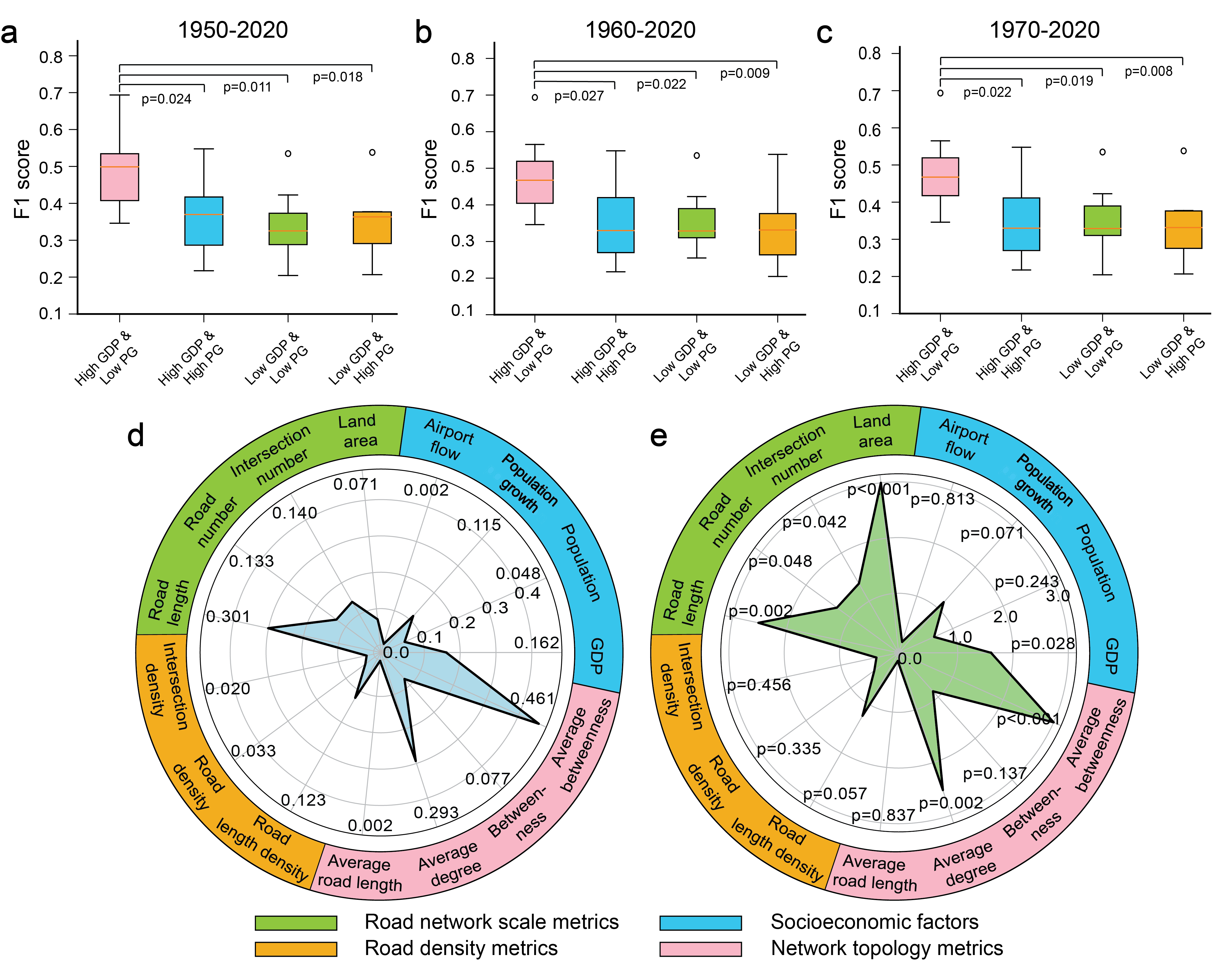}
  \label{fig:3}
\end{minipage}
\caption{\footnotesize
\textbf{Socioeconomic development revealed by the intra-city road network spatial homogeneity.} ($\textbf{a}$-$\textbf{c}$) The comparison of $F1$ scores for four types of cities defined by high and low GDP, high and low population growth (PG) from Year 1 to Year 2. The three figures differ in Year 1 (1950, 1960, and 1970) and have the same Year 2 (2020). We find that mature cities, defined as the cities in the "High GDP \& Low PG" class, have significantly higher $F1$ score than other classes, which indicates that road networks in mature cities are more predictable and homogeneous than other cities (see \textbf{Results} for more discussions). ($\textbf{d}$) $R^{2}$ values for the linear regression with a single explanatory variable when the response variable is the $F1$ score. We find average betweenness, average degree, and road length explain a large ratio of the variance of the $F1$ score. ($\textbf{e}$) $P$-values for the explanatory variables in the linear regression which is the same as ($\textbf{d}$). To better visualize small $P$-values, we replace $p$$<$0.001 with $p$$=$0.001 and draw the wind plot based on $-log(p)$. Except for the three variables mentioned in ($\textbf{d}$), we notice that the $F1$ score is sensitive to the land area.
}
\label{fig:3}
\end{figure}

\subsection*{Road network spatial homogeneity reveals socioeconomic development}
 %Our findings demonstrate that homogeneity patterns derived from the road networks enable the extraction of urban social-economic insights, which could serve as a building block in urban studies.  

Road networks serve as the skeleton of urban space, evolving over the historical process of economic development and spatial population agglomeration. To facilitate the understanding of cities, we investigate the interplay between intra-city spatial homogeneity and socioeconomic statuses such as economy, population growth, and city ages. More precisely, we define the population growth (PG) as the ratio of population in Year 2 over Year 1 to measure the increase of the population from Year 1 to Year 2. PG for all 30 cities is calculated based on the population data spanning from 1950 to 2020 (Supplementary Fig. 26). We uniformly classify each city as one of the four categories based on its GDP and PG values. For example, London (GDP=634 billion dollars, PG(1950→2020)=1.140) ranks respectively in the first half, the latter half among 30 cities in terms of GDP and PG(1950→2020), so we classify London into the “High GDP $\&$ Low PG” class. In \textbf{Figs. \ref{fig:3}$a$-$c$}, we perceive that during different time intervals, the cities in the “High GDP $\&$ Low PG” class have significantly higher prediction accuracy than the other three classes of cities ($P$-values$<$0.05 under one-sided $t$-tests). Moreover, all “High GDP $\&$ Low PG” cities (London, Paris, New York, Los Angeles, Chicago, Philadelphia, Osaka) are in developed countries and we refer to them as “mature cities”. These cities have invested much more in road network planning so that their road networks are more standardized and predictable. Moreover, the low PG in a “mature city” reveals that this city has already experienced rapid population growth and entered a stable period, therefore the forces of road network expansion are not as strong as before and consequently more predictable. We also investigate the relationships between existing network metrics (degree, betweenness) and city types, and do not find as significant results as the spatial homogeneity (Supplementary Fig. 25). Next, we study whether our new metric is associated with the age of the city. We cluster the 30 cities into 4 types (“BC”, “1-16th century”, “17-18th century”, and “19-20th century”) based on the first settlement time (Supplementary Table 5). As shown in Supplementary Fig. 28, we find the cities in “17-18th century” or “19-20th century” types have significantly higher F1 values compared to cities constructed in “BC” and “1-16th century”. This result can be expected as road networks in “younger” cities grew at the motor age when standard and united urban planning was necessary at the city level.

To further explore the interactions between spatial homogeneity and urban factors, we perform regression analysis between the spatial homogeneity and 15 measures from 4 categories: socioeconomic factors, road network scale metrics, road density metrics, and network topology metrics. We conduct linear regression with the explanatory variable as one of these fifteen factors and the response variable as the average $F1$ score for the same city. The $R^{2}$ values (\textbf{Fig. \ref{fig:3}$d$}) for the regression, the $P$-values (\textbf{Fig. \ref{fig:3}$e$}) for the coefficients of 15 factors under two-sided $t$-tests are calculated to demonstrate the statistical significance. We recognize that PG is more relevant than the population itself in relating URN spatial homogeneity ($R^{2}$: 0.115 vs 0.048, $P$-values: 0.071 vs 0.243), implying the closer connection between spatial homogeneity and population growth rate. The $F1$ score is found to be associated with GDP where higher GDP leads to higher spatial homogeneity. Moreover, the spatial homogeneity is positively related to road network scale metrics whereas road density metrics are weak predictors ($R^{2}$ $<$0.130). Note that the land area is not sufficient to solely estimate the $F1$ score ($R^{2}$=0.071) but is significantly related to $F1$ score ($P$-value$<$0.001), revealing that the urban coverage affects but does not fully determine URN network spatial homogeneity.

\subsection*{Transfer learning of road network spatial homogeneity quantitatively reveals inter-city similarity}
While socioeconomic development encodes the intra-city urban process, inter-city similarity focuses
on the interactions between different cities. 
%on the process of imitating, learning from, and transferring urban planning policies across different cities. 
These relationships present how the urban patterns emerge, spread, and evolve over the globe and are fundamental properties of human urbanization history. However, these relationships are especially hard to accurately capture and quantitatively depict due to the complex urban network patterns and the massiveness of urban data. Here, we find that inter-city similarity can be captured by our spatial homogeneity metric defined from transferring GNN models across cities.  

To compute the inter-city URN spatial homogeneity, we perform cross-city link prediction (train GNN models on city A and test on city B across 30 cities) and implement the hierarchical clustering on these cities based on their average $F1$ scores as the training sample and the testing sample. Each city is associated with 30 $F1$ scores when it serves as the training city, and the other 30 $F1$ scores when it is the testing city. For each city, we average the former 30 $F1$ scores, and the latter 30 $F1$ scores, respectively, and present the results in \textbf{Figs. \ref{fig:4}$ab$}. Furthermore, \textbf{Fig. \ref{fig:4}$c$} exhibits the City Homogeneity Transfer Matrix (CHTM) whose entry CHTM(A, B) is defined as the $F1$ score when applying the GNN trained on city A on another city B. Note that the CHTM is not symmetric: the $F1$ score obtained by training on city A and testing on city B differs from the $F1$ score when swapping city A and city B. In the CHTM, a high $F1$ score block is situated at the top-left corner (training: from Milan to Seoul; testing: from Chicago to Los Angeles). In this block, most training cities are historical European cities (Milan, London, Paris, etc.) or the US cities (Los Angeles, Chicago, New York, etc.), whereas testing cities are primarily the US cities (10 out of 13), revealing that URN patterns in the US are sufficiently covered by cities from Europe and other US cities. This result aligns with the history that the American urban planning style is inspired originally from European cities \cite{peterson2009birth}. 

The second observation is that there are four city clusters from the testing side: “USA”, “Asia(c)”, “Asia(w)”, “Europe” (\textbf{Figs. \ref{fig:4}$ac$}). In general, “USA” type cities have the highest $F1$ scores, and “Europe” type cities, all from Europe, are less predictable. The other two types with weak prediction performance are the “Asia(c)” type and the “Asia(w)” type. “Asia(c)” type cities were heavily affected by Chinese culture whereas “Asia(w)” cities had active connections with the western world. Collectively speaking, American cities were systematically planned and constructed and therefore predictable, while cities in Europe and Asia were shaped simultaneously by intricate historical and modern factors.

Another interesting observation is that setting a training city as a city from Milan to Seoul achieves higher $F1$ scores than remaining cities (\textbf{Figs. \ref{fig:4}$bc$}). We deduce that if a city contains multiple sub-structures and has a large spatial coverage, it is likely that the topological features learned from this city are applicable to other cities. In this case, we reason that these cities have excellent road network “diversity”. Most Asian cities have deficient “diversity” as they did not exert significant and enduring impacts worldwide in modern urban planning as cities in Europe and the US \cite{peng2011urbanization}. 

We now encode the city by a 16-dimension vector composed of 5 metrics (i.e., number of nodes, number of links, total link length, betweenness, and total area) plus the 11 metrics we use in the road network clustering (please find them in \textbf{Methods}). These 16 metrics include aggregate measurement (e.g., number of nodes \cite{boeing2020multi}), centrality measurement (e.g., betweenness \cite{kirkley2018betweenness}), degree measurement (e.g., average degree \cite{wang2015resilience}), circuity metrics (e.g., logarithmic circuity ($r\leq$ 0.5km) \cite{giacomin2015road}), dendricity metrics (e.g., fraction of “bridge edges” \cite{barrington2020global}), and therefore provide comprehensive descriptions of the urban road network characteristics. We use the cosine value of two vectors to calculate the inter-city similarity and demonstrate the results in \textbf{Figs. \ref{fig:4}$de$}. In general, \textbf{Fig. \ref{fig:4}$d$} exhibits two clusters (from Brussels to Shanghai; from Madrid to Chicago). 8 out of 13 cities in the second cluster are cities in the United States. Compared with CHTM in this study, this similarity matrix does not clearly separate Asian and European cities. \textbf{Fig. \ref{fig:4}$e$} fails to present interpretable patterns. In summary, the spatial homogeneity outperforms these existing network metrics in revealing inter-city road network topological similarity patterns. More details about this experiment could be found in Supplementary Section 3.1.
\begin{figure}[H]
\centering
\begin{minipage}{1.00\textwidth}
    \centering
  \includegraphics[width=16cm]{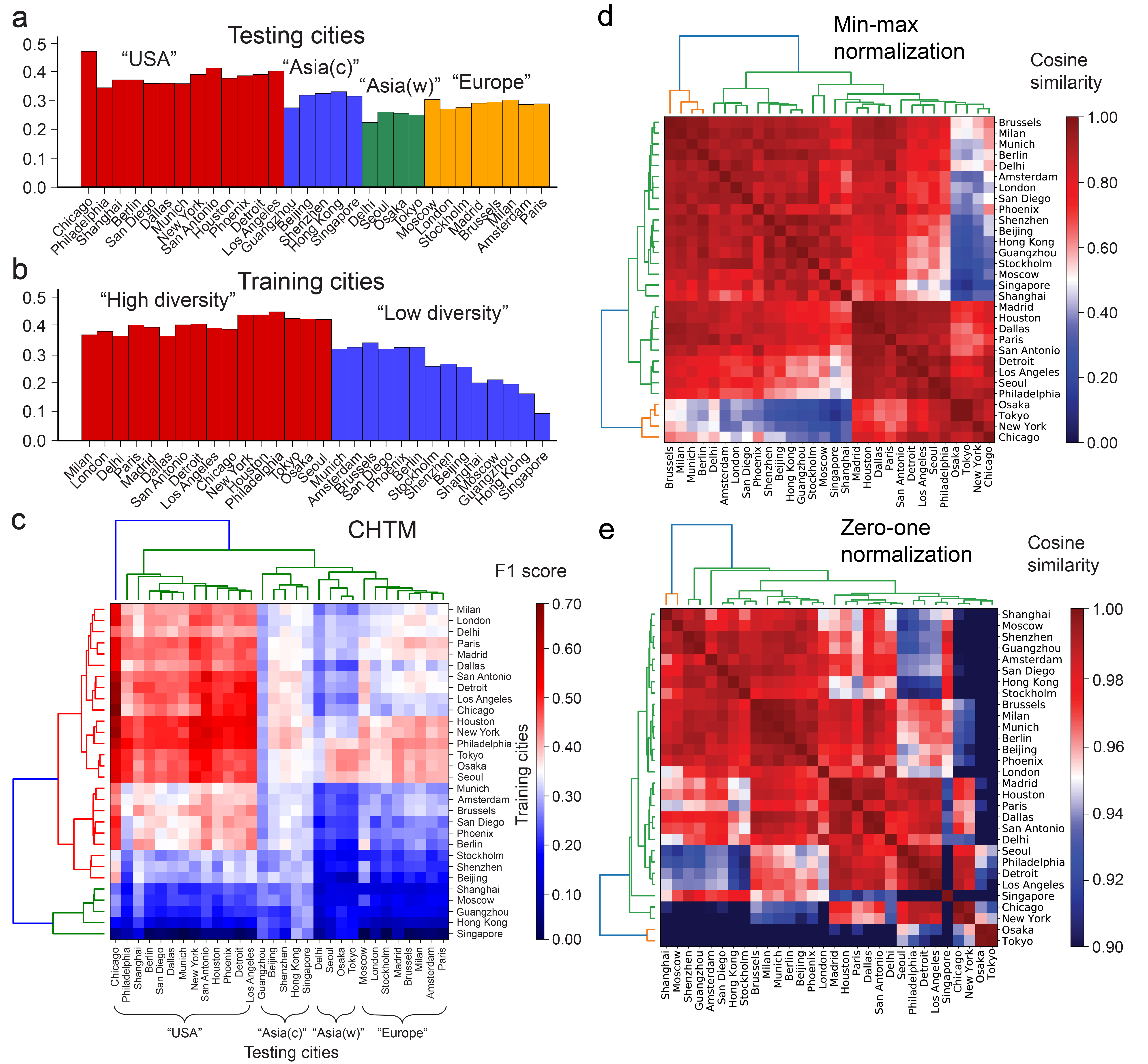}
  \label{fig:4}
\end{minipage}
\caption{\footnotesize
\textbf{Inter-city similarity revealed by the inter-city road network spatial homogeneity.} 
We train a GNN model on one city, perform the link prediction on the other cities. ($\textbf{a}$) Average $F1$ score when a city is the testing city. The USA type cities (defined from the CHTM) have higher $F1$ scores than other types of cities and are generally more predictable. ($\textbf{b}$) Average $F1$ score when a city is the training city. The cities with high diversity (defined from the CHTM) achieve higher $F1$ scores. ($\textbf{c}$) The City Homogeneity Transfer Matrix (CHTM). Each entry denotes the average $F1$ score of URNs for each training-testing city pair. CHTM reveals the URN pattern dissemination sequence from Europe to the US, and then to Asia. We cluster these cities using only the $F1$ scores, and get 4 (2) classes from the testing (training) perspectives. The clustering results are highly consistent with the continents and we refer to them as "USA type", "Asia(c) type", "Asia(w) type", "Europe type", "high diversity", and "low diversity". ($\textbf{d},\textbf{e}$) 
City similarity matrices using 16 conventional network metrics. d: We normalize the value linearly to 0 and 1 by the minimal and maximal value. e: We normalize the value by divding the value by the maximal value. 
}
\label{fig:4}
\end{figure}

\subsection*{Historical urban insights from inter-city transfer learning}
Next, we downscale the analysis and study the predictability performance of inter-city transfer learning on local subnetworks to further investigate the historical urban insights \cite{ortman2014pre}. For instance, when the models are trained on Chicago, New York and tested on Los Angeles (LA) (\textbf{Figs. \ref{fig:5}$ab$}), the south and west of downtown areas of LA achieve better prediction performance than both downtown, northern, and eastern LA, implying that southern and western LA are more consistent with Chicago and New York in terms of road network style. The early settlement in LA concentrated on the downtown area which evolved into a strong central business district (CBD) later. After the 1880s, rapid urban expansion converted the farmland into sprawled residential areas, along the direction of current Interstate-10 and Interstate-110, under a united municipal zoning ordinance which had a far-reaching influence on contemporary American cities, resulting in the high predictability of roads in these “new” areas \cite{whittemore2012zoning}. 

Contrasting results appear for a historical city such as Osaka. We find that the road networks in downtown Osaka can be well predicted by the model trained on Tokyo whereas the prediction performance for the surrounding areas is not very good (\textbf{Fig. \ref{fig:5}$c$}). It is because cultural centripetal force and the short spatial distance helped Osaka and Tokyo form a tight socioeconomic (including road network planning) connection since the period when downtown Osaka was populated. However, the populated urban land far from downtown Osaka was mostly constructed after being incorporated as a city in 1889, when both the domestic and international urban planning mode exerted influence. As a result, road networks in downtown Osaka are weakly captured by London (\textbf{Fig. \ref{fig:5}$d$}) and New York (Supplementary Fig. 27c).

The final key finding is the reclaimed land in Tokyo. We notice that URNs in the reclaimed area at the coast of Tokyo Bay are better captured by road network patterns learned from Chicago, New York (\textbf{Figs. \ref{fig:5}$ef$}), London, and Paris (Supplementary Figs. 27ef) than the broad areas in the western part of Tokyo. Administrations in Tokyo turned Tokyo Bay into usable land via landfill technologies. The “reclaimed land” marked in \textbf{Figs. \ref{fig:5}$ef$} were mostly filled after 1987 for residence, exhibition, and amusement purposes \cite{endoh2004historical}. Road networks on these artificial islands are “modern” (not follow the traditional road network styles as inland Tokyo) and share some similarities with the URNs in other international cities. 

\begin{figure}[H]
\centering
\begin{minipage}{1.00\textwidth}
    \centering
  \includegraphics[width=16cm]{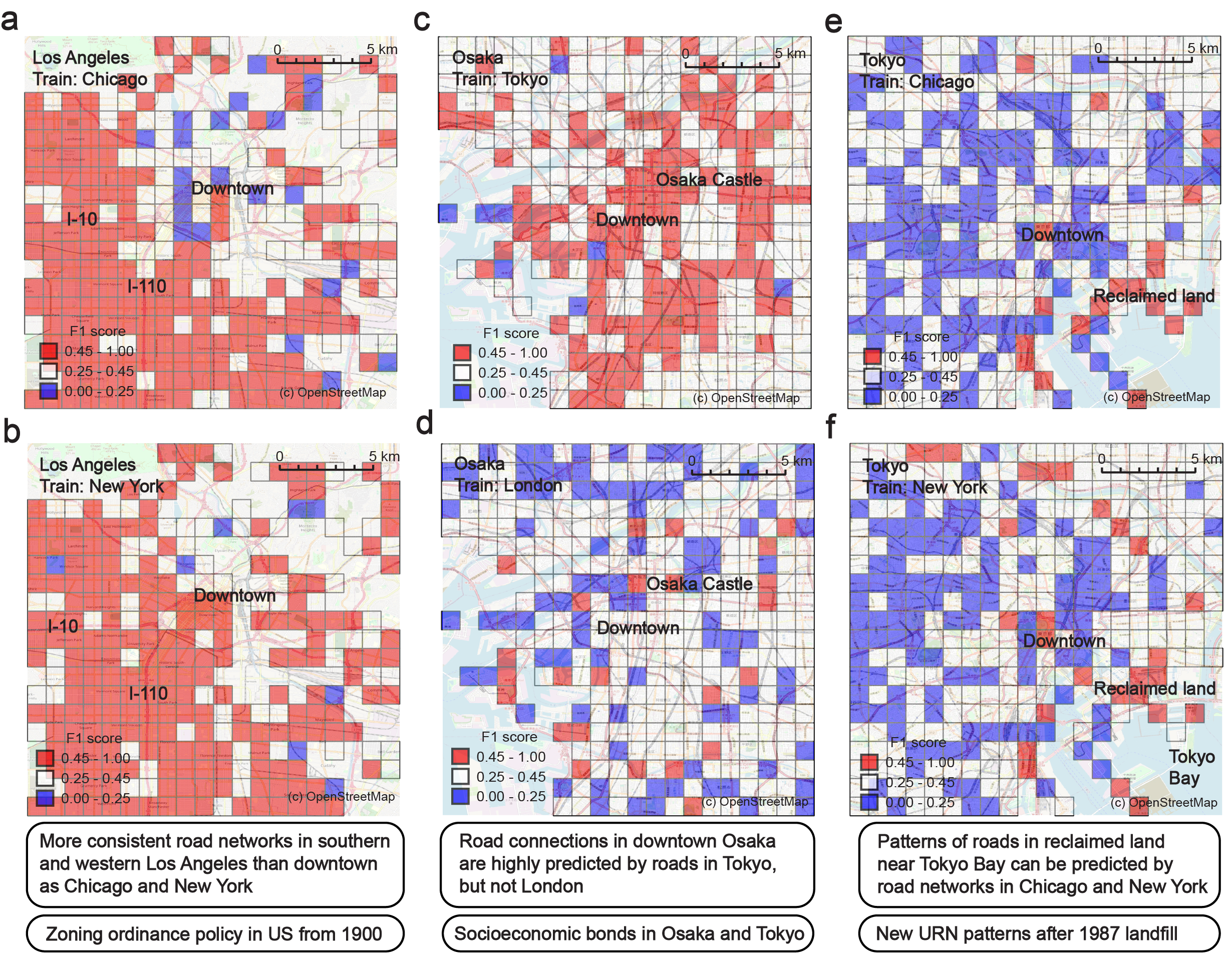}
  \label{fig:5}
\end{minipage}
\caption{\footnotesize
\textbf{Historical urban insights revealed using inter-city spatial homogeneity.} ($\textbf{a},\textbf{b}$) Spatial distribution of $F1$ scores for 1km by 1km URNs in Los Angeles (LA). Note that the red, white, and blue colors in each cell present high ($>$0.45), medium (0.25-0.45), and low ($<$0.25) $F1$ scores. The western and southern parts of LA along Interstate-10 and Interstate-110 are highly predictable. These areas were mostly urbanized after 1900 when the zoning ordinance policy in the US took effect. ($\textbf{c}, \textbf{d}$) $F1$ scores in Osaka. The URN patterns in downtown areas can be well captured by Tokyo but not London. It is because Osaka and Tokyo formed strong domestic socioeconomic communication bonds when the downtown was populated. ($\textbf{e},\textbf{f}$) $F1$ scores in Tokyo. Both Chicago and New York contain more URN patterns in the reclaimed land area in Tokyo than the western Tokyo, which is consistent with the history that Tokyo administration performed the landfill in Tokyo Bay after 1987.
}
\label{fig:5}
\end{figure}

\section*{Discussion}
Using GNN models, we propose a spatial homogeneity metric that reveals profound urban development and inter-city similarity patterns. We notice that cities with high GDP and low population growth have significantly more homogeneous road networks than other cities, revealing the hidden principle between road network structures and well-studied urban indicators used in the urban scaling law studies \cite{bettencourt2007growth,arcaute2015constructing,bettencourt2020urban, molinero2021geometry}. Compared with existing network metrics, our spatial homogeneity has the unique functionality of quantifying the bidirectional inter-city urban similarity, providing a scientific foundation for many urban insight transfer methods \cite{wei2016transfer,dai2009eigentransfer,dong2019predicting}, as well as methods to discover similarities between cities \cite{currid2010two}. 

The spatial homogeneity of URNs shares some similarities with the concept of fractal \cite{mandelbrot1982fractal,falconer1999techniques,meakin1983formation}, whose variants (multifractal, monofractal) describe the road network morphology \cite{batty1994fractal,sidqi2019comparing,molinero2021geometry,ariza2013multifractal} and characterize road network evolution patterns \cite{makse1998modeling,murcio2015multifractal}. The spatial homogeneity of URNs measures the analogy of intersection “connection” patterns between the subset and the entire network, while the fractal dimension quantifies the invariance of road network “images” under different levels of resolutions. Both metrics summarize the structure properties of URNs, however, the spatial homogeneity applies for intra- and inter-city (via the CHTM) scenarios while the fractal is used for intra-city analysis. 

This study has three limitations. First, although the GNN encodes intricate multi-hop node dependencies in road networks, the prediction performance degrades when the GNN goes deeper, suffering from vanishing gradients \cite{he2016deep} and over-smoothing \cite{chen2020measuring} issues during the model training. The prediction performance could be later improved using more advanced learning architectures and training algorithms. Second, the link prediction reliability of URNs, which is affected by node merging threshold (30m), train-test split ratio (75\%, 25\%), and link hidden percentage (20\%) (Supplementary Information), should be carefully considered. In fact, the parameter selection in road network studies is a common challenge. In the study of SNDi \cite{barrington2019global,barrington2020global}, Barrington-Leigh et al. set the node merging threshold as 20 meters after comparing it with 14 meters. In this study, we define it as 30 meters after checking the node merging results in New York and London (Supplementary Fig. 6). Comprehensive parameter sensitivity tests and a unified criterion tailored to different road networks are needed for further investigation. 
Third, we define the city region as the 20km-by-20km square area. A more optimal boundary definition should be: (1) the urban boundary from standard sources such as Metropolitan statistical areas (MSAs) \cite{berry1969metropolitan} or the Global Human Settlement Layer (GHSL) \cite{corbane2019automated} if socioeconomic data is unavailable; (2) the urban boundary defined from existing methods such as City Clustering Algorithm (CCA) \cite{rozenfeld2008laws}, commuting data-based method \cite{shen2019delineating}, cellular automata model \cite{long2016mapping}, or the percolation model \cite{cao2020quantifying} if socioeconomic data is available. In addition, we find that the boundary road segments on URNs have minor effects on the overall spatial homogeneity evaluation (Supplementary Fig. 13). Further investigation to reduce the impact of these boundary road segments is needed.

This study has many academic and practical contributions:
\begin{itemize} %[itemsep= 0 pt,topsep = 0 pt,parsep = 0pt,leftmargin = 10 pt]
\item \textbf{Missing data completion (Humanitarian and mapping assistance)}: In road network data-sparse regions, managers employ the GNN models to predict missing roads, and then verify their existence after comparing them with ground truth data such as satellite images.

\item \textbf{Infrastructure equity evaluation (Urban science and policy)}: In rapidly urbanizing cities, regional and social scientists leverage this metric to measure road network spatial homogeneity in different regions, and make equitable infrastructure and facility policy decisions \cite{hanson2004geography}.

\item \textbf{Quantitative urban analysis (Urban archaeology)}: Using the inter-city similarity results, urban archaeologists can better track the historical evolution of global road networks \cite{zischg2019century} and the socioeconomic environment in a quantitative manner.

\item \textbf{Transfer learning benchmarking (Urban computing)}: Urban transfer learning methods \cite{dai2009eigentransfer,dong2019predicting,yabe2020unsupervised} transfer knowledge across cities, but ignore whether the transferring performance is universally adequate for different cities. Owing to the strong dependence of socioeconomic activities with URNs, measuring the spatial homogeneity of URNs across cities could assist the evaluation of inter-city transferability. 

\item \textbf{Cascading failure analysis (Network science)}: The cascading failure evolves from the local to multi-hop neighbors and global scales in the network \cite{zhao2016spatio}. The spatial homogeneity may be applied to understanding the cascading dynamics of failures, given that the metric captures multi-hop node relationships in the network.
\end{itemize}

In summary, the spatial homogeneity metric quantifies the similarity between the subsection and the entire network, and provides a wide variety of multidisciplinary contributions. In the future, high-granularity road networks and urban traffic data (such as road hierarchy, traffic flow) would promote the quantification of the traffic-network spatial homogeneity. In addition, user-friendly spatial homogeneity computing packages are also favorable to increase the machine learning implementation efficiency in practices. This study advances our understanding of complex road networks building on seminal work \cite{li2015percolation,saberi2020simple,ccolak2016understanding,loder2019understanding,zeng2020multiple} in recent years.

\section*{Methods}
\subsection*{Datasets}
We utilize the OSMnx \cite{boeing2017osmnx} to download the road network data from OpenStreetMap (OSM), which is widely used for academic purposes \cite{kirkley2018betweenness,ganin2017resilience,boeing2020multi}. We chose 30 major global cities (see Supplementary Information for more details) as samples to accommodate diverse cities in terms of age (ancient and modern cities), regions (cities in the US, Europe, and Asia), locations (coastal and inland cities), and development statuses (cities in developed and developing countries). The intuition of selecting mainly metropolitan cities is that road networks in metropolitan cities are representative of the road network planning paradigm in a specific region. 

For each city, we define the study area as a 0.2 degrees ($\sim$20km$\times$20km) grid area with 400 URNs with the size of 1km-by-1km (Supplementary Fig. 1). Note that grid cells have been widely used in existing urban studies \cite{louail2014mobile,thompson2020global,abitbol2020interpretable,barrington2020global}. Despite the discrepancy of research objectives, the fundamental principle is that the grid cell size should fit the specific task. In our study, we find the 1km-by-1km grid cell contains a sufficient number of intersections and road segments for us to train the link prediction model, and thus determine the grid cell size as 1km-by-1km. In particular, we first build the grid around the city centroid point with the spatial distance as 0.01 degrees in the longitude and latitude directions. Next, we extract the road network data using OSMnx by calling the $G = ox.graph\_from\_point(location, distance, distance\_type=bbox,  network\_type=drive)$ function. Here, the function input variable $location$ represents the longitude and latitude of the centroid of the extracted area, and the function input variable $distance$ represents half of the side length of the square area surrounding the centroid of the extracted area. Next, we study whether our definition of the URN as the grid cell affects the overall F1 scores. In particular, we shift the centroid of the 20km-by-20km grid for each city by half of the unit so that most boundary road segments in the previous setting are included within the URN after the centroid shifting. We re-train the machine-learning model and find  (1) that both the median and variance of $F1$ score distribution for 30 cities changes slightly after the centroid shifting; (2) the spatial homogeneity order (from Chicago to Moscow) keeps stable despite some local changes (Supplementary Fig. 13).

We visualize 20km-by-20km city regions for all 30 cities in Supplementary Figs. 1,2,3,4. It can be seen that both the central and peripheral regions of the city have been included. We also compute spatial homogeneity values for 6 cities when the regions are defined as 30km-by-30km, and find that their $F1$ score distributions are highly close to the 20km-by-20km case (Supplementary Figs. 10,11,12). More importantly, the urban knowledge “the road networks in Chicago and Singapore have the highest and lowest spatial homogeneity, respectively” can be invariantly concluded for both 20km-by-20km and 30km-by-30km cases. The alternative city boundary definitions can be Metropolitan Statistical Areas (MSAs)\cite{giacomin2015road}, Larger Urban Zones (LUZs), or EU Urban Atlas (EUAs)\cite{ua2012}. Note that MSAs (LUZs, EUAs) are the datasets for only the American (European, European) cities respectively. Owing to the lack of standard commuting data in all 30 cities, in this study we do not refer to urban functional delineation \cite{shen2019delineating,khiali2019combining, rozenblat2020extending,ma2020functional}, which is an alternative to defining urban boundaries. Given that our objective includes comparing and transferring the spatial homogeneity across cities, we decide the consistent sampling sizes that include the urban cores and surrounding areas to avoid the exterior impact from variant urban sizes. We set the training and testing sample ratio as 3:1 (Supplementary Fig. 5). To deal with the “duplicate node problem”, we implement a node merging algorithm and verify the merging performance in New York and London (Supplementary Fig. 6). In addition, we collect the social indicator data (e.g., GDP, population) used in the association analysis (see Supplementary Information for more details).

\subsection*{GNN model}
We predict the missing links based on the URN topology. It is easy to extend our model by adding known socioeconomic factors (e.g., land use, demographic statistics) as the node and link features, or predict other types of information (e.g., the number of road lanes, traffic flows) by formulating the task as a multi-class classification problem. Our machine-learning task is the link prediction on URNs, where we sample 80\% of links and predict the presence of the rest 20\% (Supplementary Information). The 11,790 URN data from 30 cities are fed into an R-GCN model to generate a low dimensional representation $h_{i}$ for each node $i$, which is a non-linear aggregation of its neighbors’ representation $h_{j}$ at the previous layer:
{\setlength\abovedisplayskip{8pt}
\setlength\belowdisplayskip{8pt}
\begin{equation}\centering
    h_{i}^{l+1} = tanh(\sum_{r \in R}\sum_{j \in N_{i}^{r}}\frac{1}{c_{i}^{r}}W_{r}^{l}h_{j}^{l}+W_{0}^{l}h_{i}^{l}),
\end{equation}}
where $l+1$ and $l$ are the layer numbers, $R$ is the relation set (Supplementary Section 2.3), $N_{i}^{r}$ is the neighborhood node set of node $i$ under the relation $r$, $W_{r}^{l}$ and $W_{0}^{l}$ are learnable parameter matrices.
Next, we adopt a bilinear projection function, named \textit{DistMult} \cite{yang2014embedding} to transform the representation of a pair of nodes into their connecting strength:
{\setlength\abovedisplayskip{8pt}
\setlength\belowdisplayskip{8pt}
\begin{equation}
    s(i,j)=h_{i}^{T}W_{d}h_{j}
\end{equation}}
and then use the sigmoid function to transform the range of scoring strength to (0,1) using $\sigma(s(i,j)) = \frac{1}{1+e^{-s(i,j)}}$.
If the sigmoid connecting strength for a pair of nodes exceeds the threshold $\delta$ ($\sigma(s(i,j))>\delta$), then we predict that the link $(i,j)$ exists. To further fit our model to this task, we also implement R-GCN*, which appends two decoding improvements to the original R-GCN, to avoid “acute angles” and “intersecting links” in the prediction (Supplementary Fig. \textcolor{red}{7}). To compare the model performance, we also implement five other GNN models (node2vec \cite{grover2016node2vec}, struc2vec \cite{ribeiro2017struc2vec}, GraphSAGE \cite{kipf2016semi}, spectral GCN \cite{hamilton2017inductive}, graph attention network (GAT) \cite{velivckovic2017graph}, Supplementary Information) to benchmark the prediction. Using Adam optimizer \cite{kingma2014adam} minimizing the cross-entropy loss, we jointly train parameters in R-GCN and DistMult in an end-to-end manner. In the intra-city prediction, we train separate models for different cities whereas in the inter-city prediction, we train on city $A$ and test on city $B$. 

\subsection*{Model performance}
As “link prediction” is a highly imbalanced machine-learning task (more negative data than positive data), we evaluate the prediction performance using both the area under the precision-recall curve (AUPRC) and $F1$ score which is a weighted average of precision and recall. Overall, the R-GCN* achieves the prediction accuracy with AUPRC=0.28 and $F1$ score=0.42 and outperforms other models, demonstrating the power of R-GCN* in capturing the link existence patterns of a URN. The performance illustrates large variance among different cities with a maximum of 0.690 (Chicago) and a minimum of 0.199 (Stockholm). Since all models provide the connecting strength for each pair of nodes, we define the sigmoid connecting strength threshold as 0.61 by maximizing the $F1$ score in the best model, R-GCN*. If the sigmoid connecting strength for two nodes is at least 0.61, then we predict they are joined by a road segment in the real world, because their structural roles are similar in the road network. Complete receiver operator curves and precision recall curves can be found in Supplementary Figs. 8,9.

\subsection*{Spatial homogeneity quantification}
The spatial homogeneity describes the similarity between the local and the entire network. A network with high spatial homogeneity tends to be predictable because the link prediction utilizes the global road network information to predict the links in the local road network. Hence, we use the prediction performance metric, $F1$ score, to quantify the spatial homogeneity:
\begin{itemize}
%[itemsep= 0 pt,topsep = 0 pt,parsep = 0pt,leftmargin = 10 pt]
\item for each road network unit, we define the \textbf{spatial homogeneity} as \textbf{the $\mathbf{F1}$ score} under the link prediction. \item for each city, we define the \textbf{city spatial homogeneity} as \textbf{the average $\mathbf{F1}$ score} of all road network units.
\end{itemize}

\subsection*{Association analysis}
We perform the association analysis between the spatial homogeneity and network type, network irregularity (Supplementary Information). Specially, we follow Barrington-Leigh and Millard-Ball \cite{barrington2019global,barrington2020global} to obtain the URN type and compute the NI. We compute 11 measurements (i.e., average degree, fractions of "degree=1", "degree=2", "degree=3", "degree$\geq$4" nodes, logarithmic circuity ($r\leq 0.5km$), logarithmic circuity ($r>0.5km$), fraction of "bridge edges" length, fraction of "bridge edges" number, fraction of "dead-end edges" length, and fraction of "dead-end edges" number) from 3 aspects, i.e., node degree, circuity, and dendricity (please find the description of these 11 measurements in Supplementary Tables 3,4 and the explanation in Supplementary Section 3.2). Next, we try distinct numbers $k$ of clusters and notice under the case of $k=4$, partitions of URNs have the best interpretability (Supplementary Fig. 18). It should be noted that a similar study that utilizes the urban image data to cluster the global cities \cite{thompson2020global} might be useful to extend our spatial homogeneity definition idea from URNs to urban images. A detailed description can be found in Supplementary Fig. 16. Despite the input data difference, we find the Adjusted Rank Index (ARI) between our city clusters from the testing data in the CHTM (i.e., "USA", "Asia(c)", "Asia(w)", and "Europe") and their city clusters (i.e., "Motor City", "High Transit", "Intense", "Sparse", "Chequerboard", and "Informal") is 0.535, demonstrating certain consistency among the road network patterns among those cities. Please find the city cluster results in Supplementary Table 2. Note that the urban image data they use contains information about urban transit so that it can be reflected in their cluster results. In addition, we perform the principal component analysis on these vectors (Supplementary Table 4), using the PC1 to define the NI, which represents the sprawl of a URN \cite{barrington2020global}. 
\subsection*{Hyperparameters}
In this link prediction task, the number of negative samples (nonexistent links) is much larger than the positive samples (links). To prohibit the GNN model from focusing too much on the negative samples, we sample a subset of negative samples to make the ratio between positive and negative samples 1:5. The number of layers, epochs, learning rate, and neurons in the GNN are hyperparameters in the model. To determine these hyperparameters, we perform the standard five-fold cross validation. In particular, we set the number of epochs as 10, the learning rate as 0.001, the number of layers as 3, and the number of neurons as 50.

\section*{Data and code availability}
We use the publicly available road network data from OpenStreetMap (https://www.openstreetmap.org/) via the OSMnx Python package  (https://github.com/gboeing/osmnx). We also use images from Google Maps (https://www.google.com/maps) to validate the node merging results. These images are also available to the public. The population data we use comes from https://worldpopulationreview.com/, which is a visualization platform of the open datasets owned by the United Nations. The airport flow data is from https://www.panynj.gov/airports/en/statistics-general-info.html owned by The Port Authority of New York and New Jersey and is also publicly available. Both the road network data, socioeconomic data, and source codes for training, testing results are available at the online data warehouse:\;\;\;\href{https://github.com/jiang719/road-network-predictability.git}{https://github.com/jiang719/road-network-predictability.git}. 

\section*{Correspondence} 
Correspondence should be addressed to S.V.U. (email: sukkusur@purdue.edu) and J.M. (email: 
majianzhu@pku.edu.cn).
\section*{Author contributions} J.X., T.Y., S.V.U., and J.M. proposed the question; J.X., N.J., S.L., Q.P., and J.M. designed the research; N.J. trained and tested the GNN models; S.L. and Q.P. performed the intra-city analysis; N.J. and J.X. conducted the inter-city analysis; J.X. and J.M. drew the figures; J.X., T.Y., S.V.U., and J.M. wrote the paper.

\section*{Competing interests} The authors declare no conflict of interest.

\section*{Acknowledgements} We thank the discussion with Prof. Suresh Rao from Purdue University about the comparison between the spatial homogeneity metric and existing network metrics.

\bibliography{sample}

\begin{thebibliography}{100}
\urlstyle{rm}
\expandafter\ifx\csname url\endcsname\relax
  \def\url#1{\texttt{#1}}\fi
\expandafter\ifx\csname urlprefix\endcsname\relax\def\urlprefix{URL }\fi
\expandafter\ifx\csname doiprefix\endcsname\relax\def\doiprefix{DOI: }\fi
\providecommand{\bibinfo}[2]{#2}
\providecommand{\eprint}[2][]{\url{#2}}

\bibitem{sun2013understanding}
\bibinfo{author}{Sun, L.}, \bibinfo{author}{Axhausen, K.~W.},
  \bibinfo{author}{Lee, D.-H.} \& \bibinfo{author}{Huang, X.}
\newblock \bibinfo{journal}{\bibinfo{title}{Understanding metropolitan patterns
  of daily encounters}}.
\newblock {\emph{\JournalTitle{Proceedings of the National Academy of
  Sciences}}} \textbf{\bibinfo{volume}{110}}, \bibinfo{pages}{13774--13779}
  (\bibinfo{year}{2013}).

\bibitem{roth2011structure}
\bibinfo{author}{Roth, C.}, \bibinfo{author}{Kang, S.~M.},
  \bibinfo{author}{Batty, M.} \& \bibinfo{author}{Barth{\'e}lemy, M.}
\newblock \bibinfo{journal}{\bibinfo{title}{Structure of urban movements:
  polycentric activity and entangled hierarchical flows}}.
\newblock {\emph{\JournalTitle{PloS one}}} \textbf{\bibinfo{volume}{6}},
  \bibinfo{pages}{e15923} (\bibinfo{year}{2011}).

\bibitem{steadieseifi2014multimodal}
\bibinfo{author}{SteadieSeifi, M.}, \bibinfo{author}{Dellaert, N.~P.},
  \bibinfo{author}{Nuijten, W.}, \bibinfo{author}{Van~Woensel, T.} \&
  \bibinfo{author}{Raoufi, R.}
\newblock \bibinfo{journal}{\bibinfo{title}{Multimodal freight transportation
  planning: A literature review}}.
\newblock {\emph{\JournalTitle{European journal of operational research}}}
  \textbf{\bibinfo{volume}{233}}, \bibinfo{pages}{1--15}
  (\bibinfo{year}{2014}).

\bibitem{bettencourt2007growth}
\bibinfo{author}{Bettencourt, L.~M.}, \bibinfo{author}{Lobo, J.},
  \bibinfo{author}{Helbing, D.}, \bibinfo{author}{K{\"u}hnert, C.} \&
  \bibinfo{author}{West, G.~B.}
\newblock \bibinfo{journal}{\bibinfo{title}{Growth, innovation, scaling, and
  the pace of life in cities}}.
\newblock {\emph{\JournalTitle{Proceedings of the national academy of
  sciences}}} \textbf{\bibinfo{volume}{104}}, \bibinfo{pages}{7301--7306}
  (\bibinfo{year}{2007}).

\bibitem{arcaute2015constructing}
\bibinfo{author}{Arcaute, E.} \emph{et~al.}
\newblock \bibinfo{journal}{\bibinfo{title}{Constructing cities, deconstructing
  scaling laws}}.
\newblock {\emph{\JournalTitle{Journal of the royal society interface}}}
  \textbf{\bibinfo{volume}{12}}, \bibinfo{pages}{20140745}
  (\bibinfo{year}{2015}).

\bibitem{xu2020deconstructing}
\bibinfo{author}{Xu, Y.}, \bibinfo{author}{Olmos, L.~E.},
  \bibinfo{author}{Abbar, S.} \& \bibinfo{author}{Gonz{\'a}lez, M.~C.}
\newblock \bibinfo{journal}{\bibinfo{title}{Deconstructing laws of
  accessibility and facility distribution in cities}}.
\newblock {\emph{\JournalTitle{Science advances}}}
  \textbf{\bibinfo{volume}{6}}, \bibinfo{pages}{eabb4112}
  (\bibinfo{year}{2020}).

\bibitem{snellen2002urban}
\bibinfo{author}{Snellen, D.}, \bibinfo{author}{Borgers, A.} \&
  \bibinfo{author}{Timmermans, H.}
\newblock \bibinfo{journal}{\bibinfo{title}{Urban form, road network type, and
  mode choice for frequently conducted activities: a multilevel analysis using
  quasi-experimental design data}}.
\newblock {\emph{\JournalTitle{Environment and Planning A}}}
  \textbf{\bibinfo{volume}{34}}, \bibinfo{pages}{1207--1220}
  (\bibinfo{year}{2002}).

\bibitem{wang2012understanding}
\bibinfo{author}{Wang, P.}, \bibinfo{author}{Hunter, T.},
  \bibinfo{author}{Bayen, A.~M.}, \bibinfo{author}{Schechtner, K.} \&
  \bibinfo{author}{Gonz{\'a}lez, M.~C.}
\newblock \bibinfo{journal}{\bibinfo{title}{Understanding road usage patterns
  in urban areas}}.
\newblock {\emph{\JournalTitle{Scientific reports}}}
  \textbf{\bibinfo{volume}{2}}, \bibinfo{pages}{1--6} (\bibinfo{year}{2012}).

\bibitem{zhan2017dynamics}
\bibinfo{author}{Zhan, X.}, \bibinfo{author}{Ukkusuri, S.~V.} \&
  \bibinfo{author}{Rao, P. S.~C.}
\newblock \bibinfo{journal}{\bibinfo{title}{Dynamics of functional failures and
  recovery in complex road networks}}.
\newblock {\emph{\JournalTitle{Physical Review E}}}
  \textbf{\bibinfo{volume}{96}}, \bibinfo{pages}{052301}
  (\bibinfo{year}{2017}).

\bibitem{li2015percolation}
\bibinfo{author}{Li, D.} \emph{et~al.}
\newblock \bibinfo{journal}{\bibinfo{title}{Percolation transition in dynamical
  traffic network with evolving critical bottlenecks}}.
\newblock {\emph{\JournalTitle{Proceedings of the National Academy of
  Sciences}}} \textbf{\bibinfo{volume}{112}}, \bibinfo{pages}{669--672}
  (\bibinfo{year}{2015}).

\bibitem{saberi2020simple}
\bibinfo{author}{Saberi, M.} \emph{et~al.}
\newblock \bibinfo{journal}{\bibinfo{title}{A simple contagion process
  describes spreading of traffic jams in urban networks}}.
\newblock {\emph{\JournalTitle{Nature communications}}}
  \textbf{\bibinfo{volume}{11}}, \bibinfo{pages}{1--9} (\bibinfo{year}{2020}).

\bibitem{ccolak2016understanding}
\bibinfo{author}{{\c{C}}olak, S.}, \bibinfo{author}{Lima, A.} \&
  \bibinfo{author}{Gonz{\'a}lez, M.~C.}
\newblock \bibinfo{journal}{\bibinfo{title}{Understanding congested travel in
  urban areas}}.
\newblock {\emph{\JournalTitle{Nature communications}}}
  \textbf{\bibinfo{volume}{7}}, \bibinfo{pages}{1--8} (\bibinfo{year}{2016}).

\bibitem{zhang2019scale}
\bibinfo{author}{Zhang, L.} \emph{et~al.}
\newblock \bibinfo{journal}{\bibinfo{title}{Scale-free resilience of real
  traffic jams}}.
\newblock {\emph{\JournalTitle{Proceedings of the National Academy of
  Sciences}}} \textbf{\bibinfo{volume}{116}}, \bibinfo{pages}{8673--8678}
  (\bibinfo{year}{2019}).

\bibitem{foley2005global}
\bibinfo{author}{Foley, J.~A.} \emph{et~al.}
\newblock \bibinfo{journal}{\bibinfo{title}{Global consequences of land use}}.
\newblock {\emph{\JournalTitle{science}}} \textbf{\bibinfo{volume}{309}},
  \bibinfo{pages}{570--574} (\bibinfo{year}{2005}).

\bibitem{strano2017scaling}
\bibinfo{author}{Strano, E.} \emph{et~al.}
\newblock \bibinfo{journal}{\bibinfo{title}{The scaling structure of the global
  road network}}.
\newblock {\emph{\JournalTitle{Royal Society open science}}}
  \textbf{\bibinfo{volume}{4}}, \bibinfo{pages}{170590} (\bibinfo{year}{2017}).

\bibitem{molinero2017angular}
\bibinfo{author}{Molinero, C.}, \bibinfo{author}{Murcio, R.} \&
  \bibinfo{author}{Arcaute, E.}
\newblock \bibinfo{journal}{\bibinfo{title}{The angular nature of road
  networks}}.
\newblock {\emph{\JournalTitle{Scientific reports}}}
  \textbf{\bibinfo{volume}{7}}, \bibinfo{pages}{1--11} (\bibinfo{year}{2017}).

\bibitem{kalapala2006scale}
\bibinfo{author}{Kalapala, V.}, \bibinfo{author}{Sanwalani, V.},
  \bibinfo{author}{Clauset, A.} \& \bibinfo{author}{Moore, C.}
\newblock \bibinfo{journal}{\bibinfo{title}{Scale invariance in road
  networks}}.
\newblock {\emph{\JournalTitle{Physical Review E}}}
  \textbf{\bibinfo{volume}{73}}, \bibinfo{pages}{026130}
  (\bibinfo{year}{2006}).

\bibitem{porta2006network}
\bibinfo{author}{Porta, S.}, \bibinfo{author}{Crucitti, P.} \&
  \bibinfo{author}{Latora, V.}
\newblock \bibinfo{journal}{\bibinfo{title}{The network analysis of urban
  streets: A dual approach}}.
\newblock {\emph{\JournalTitle{Physica A: Statistical Mechanics and its
  Applications}}} \textbf{\bibinfo{volume}{369}}, \bibinfo{pages}{853--866}
  (\bibinfo{year}{2006}).

\bibitem{crucitti2006centrality}
\bibinfo{author}{Crucitti, P.}, \bibinfo{author}{Latora, V.} \&
  \bibinfo{author}{Porta, S.}
\newblock \bibinfo{journal}{\bibinfo{title}{Centrality measures in spatial
  networks of urban streets}}.
\newblock {\emph{\JournalTitle{Physical Review E}}}
  \textbf{\bibinfo{volume}{73}}, \bibinfo{pages}{036125}
  (\bibinfo{year}{2006}).

\bibitem{kirkley2018betweenness}
\bibinfo{author}{Kirkley, A.}, \bibinfo{author}{Barbosa, H.},
  \bibinfo{author}{Barthelemy, M.} \& \bibinfo{author}{Ghoshal, G.}
\newblock \bibinfo{journal}{\bibinfo{title}{From the betweenness centrality in
  street networks to structural invariants in random planar graphs}}.
\newblock {\emph{\JournalTitle{Nature communications}}}
  \textbf{\bibinfo{volume}{9}}, \bibinfo{pages}{1--12} (\bibinfo{year}{2018}).

\bibitem{jiang2004topological}
\bibinfo{author}{Jiang, B.} \& \bibinfo{author}{Claramunt, C.}
\newblock \bibinfo{journal}{\bibinfo{title}{Topological analysis of urban
  street networks}}.
\newblock {\emph{\JournalTitle{Environment and Planning B: Planning and
  design}}} \textbf{\bibinfo{volume}{31}}, \bibinfo{pages}{151--162}
  (\bibinfo{year}{2004}).

\bibitem{louf2014typology}
\bibinfo{author}{Louf, R.} \& \bibinfo{author}{Barthelemy, M.}
\newblock \bibinfo{journal}{\bibinfo{title}{A typology of street patterns}}.
\newblock {\emph{\JournalTitle{Journal of The Royal Society Interface}}}
  \textbf{\bibinfo{volume}{11}}, \bibinfo{pages}{20140924}
  (\bibinfo{year}{2014}).

\bibitem{lee2017morphology}
\bibinfo{author}{Lee, M.}, \bibinfo{author}{Barbosa, H.},
  \bibinfo{author}{Youn, H.}, \bibinfo{author}{Holme, P.} \&
  \bibinfo{author}{Ghoshal, G.}
\newblock \bibinfo{journal}{\bibinfo{title}{Morphology of travel routes and the
  organization of cities}}.
\newblock {\emph{\JournalTitle{Nature communications}}}
  \textbf{\bibinfo{volume}{8}}, \bibinfo{pages}{1--10} (\bibinfo{year}{2017}).

\bibitem{masucci2015problem}
\bibinfo{author}{Masucci, A.~P.}, \bibinfo{author}{Arcaute, E.},
  \bibinfo{author}{Hatna, E.}, \bibinfo{author}{Stanilov, K.} \&
  \bibinfo{author}{Batty, M.}
\newblock \bibinfo{journal}{\bibinfo{title}{On the problem of boundaries and
  scaling for urban street networks}}.
\newblock {\emph{\JournalTitle{Journal of the Royal Society Interface}}}
  \textbf{\bibinfo{volume}{12}}, \bibinfo{pages}{20150763}
  (\bibinfo{year}{2015}).

\bibitem{lammer2006scaling}
\bibinfo{author}{L{\"a}mmer, S.}, \bibinfo{author}{Gehlsen, B.} \&
  \bibinfo{author}{Helbing, D.}
\newblock \bibinfo{journal}{\bibinfo{title}{Scaling laws in the spatial
  structure of urban road networks}}.
\newblock {\emph{\JournalTitle{Physica A: Statistical Mechanics and its
  Applications}}} \textbf{\bibinfo{volume}{363}}, \bibinfo{pages}{89--95}
  (\bibinfo{year}{2006}).

\bibitem{depersin2018global}
\bibinfo{author}{Depersin, J.} \& \bibinfo{author}{Barthelemy, M.}
\newblock \bibinfo{journal}{\bibinfo{title}{From global scaling to the dynamics
  of individual cities}}.
\newblock {\emph{\JournalTitle{Proceedings of the National Academy of
  Sciences}}} \textbf{\bibinfo{volume}{115}}, \bibinfo{pages}{2317--2322}
  (\bibinfo{year}{2018}).

\bibitem{thadakamalla2005search}
\bibinfo{author}{Thadakamalla, H.~P.}, \bibinfo{author}{Albert, R.} \&
  \bibinfo{author}{Kumara, S.~R.}
\newblock \bibinfo{journal}{\bibinfo{title}{Search in weighted complex
  networks}}.
\newblock {\emph{\JournalTitle{Physical Review E}}}
  \textbf{\bibinfo{volume}{72}}, \bibinfo{pages}{066128}
  (\bibinfo{year}{2005}).

\bibitem{jeong2007low}
\bibinfo{author}{Jeong, J.} \& \bibinfo{author}{Berman, P.}
\newblock \bibinfo{journal}{\bibinfo{title}{Low-cost search in scale-free
  networks}}.
\newblock {\emph{\JournalTitle{Physical Review E}}}
  \textbf{\bibinfo{volume}{75}}, \bibinfo{pages}{036104}
  (\bibinfo{year}{2007}).

\bibitem{ahmadzai2019assessment}
\bibinfo{author}{Ahmadzai, F.}, \bibinfo{author}{Rao, K.~L.} \&
  \bibinfo{author}{Ulfat, S.}
\newblock \bibinfo{journal}{\bibinfo{title}{Assessment and modelling of urban
  road networks using integrated graph of natural road network (a gis-based
  approach)}}.
\newblock {\emph{\JournalTitle{Journal of Urban Management}}}
  \textbf{\bibinfo{volume}{8}}, \bibinfo{pages}{109--125}
  (\bibinfo{year}{2019}).

\bibitem{nigam2021local}
\bibinfo{author}{Nigam, R.}, \bibinfo{author}{Sharma, D.~K.},
  \bibinfo{author}{Jain, S.} \& \bibinfo{author}{Srivastava, G.}
\newblock \bibinfo{journal}{\bibinfo{title}{A local betweenness centrality
  based forwarding technique for social opportunistic iot networks}}.
\newblock {\emph{\JournalTitle{Mobile Networks and Applications}}}
  \bibinfo{pages}{1--16} (\bibinfo{year}{2021}).

\bibitem{porta2012street}
\bibinfo{author}{Porta, S.} \emph{et~al.}
\newblock \bibinfo{journal}{\bibinfo{title}{Street centrality and the location
  of economic activities in barcelona}}.
\newblock {\emph{\JournalTitle{Urban studies}}} \textbf{\bibinfo{volume}{49}},
  \bibinfo{pages}{1471--1488} (\bibinfo{year}{2012}).

\bibitem{mahyar2019compressive}
\bibinfo{author}{Mahyar, H.}, \bibinfo{author}{Hasheminezhad, R.} \&
  \bibinfo{author}{Stanley, H.~E.}
\newblock \bibinfo{journal}{\bibinfo{title}{Compressive closeness in
  networks}}.
\newblock {\emph{\JournalTitle{Applied Network Science}}}
  \textbf{\bibinfo{volume}{4}}, \bibinfo{pages}{1--21} (\bibinfo{year}{2019}).

\bibitem{schneider2013unravelling}
\bibinfo{author}{Schneider, C.~M.}, \bibinfo{author}{Belik, V.},
  \bibinfo{author}{Couronn{\'e}, T.}, \bibinfo{author}{Smoreda, Z.} \&
  \bibinfo{author}{Gonz{\'a}lez, M.~C.}
\newblock \bibinfo{journal}{\bibinfo{title}{Unravelling daily human mobility
  motifs}}.
\newblock {\emph{\JournalTitle{Journal of The Royal Society Interface}}}
  \textbf{\bibinfo{volume}{10}}, \bibinfo{pages}{20130246}
  (\bibinfo{year}{2013}).

\bibitem{dey2019network}
\bibinfo{author}{Dey, A.~K.}, \bibinfo{author}{Gel, Y.~R.} \&
  \bibinfo{author}{Poor, H.~V.}
\newblock \bibinfo{journal}{\bibinfo{title}{What network motifs tell us about
  resilience and reliability of complex networks}}.
\newblock {\emph{\JournalTitle{Proceedings of the National Academy of
  Sciences}}} \textbf{\bibinfo{volume}{116}}, \bibinfo{pages}{19368--19373}
  (\bibinfo{year}{2019}).

\bibitem{benson2018simplicial}
\bibinfo{author}{Benson, A.~R.}, \bibinfo{author}{Abebe, R.},
  \bibinfo{author}{Schaub, M.~T.}, \bibinfo{author}{Jadbabaie, A.} \&
  \bibinfo{author}{Kleinberg, J.}
\newblock \bibinfo{journal}{\bibinfo{title}{Simplicial closure and higher-order
  link prediction}}.
\newblock {\emph{\JournalTitle{Proceedings of the National Academy of
  Sciences}}} \textbf{\bibinfo{volume}{115}}, \bibinfo{pages}{E11221--E11230}
  (\bibinfo{year}{2018}).

\bibitem{chandra2000does}
\bibinfo{author}{Chandra, A.} \& \bibinfo{author}{Thompson, E.}
\newblock \bibinfo{journal}{\bibinfo{title}{Does public infrastructure affect
  economic activity?: Evidence from the rural interstate highway system}}.
\newblock {\emph{\JournalTitle{Regional Science and Urban Economics}}}
  \textbf{\bibinfo{volume}{30}}, \bibinfo{pages}{457--490}
  (\bibinfo{year}{2000}).

\bibitem{molinero2021geometry}
\bibinfo{author}{Molinero, C.} \& \bibinfo{author}{Thurner, S.}
\newblock \bibinfo{journal}{\bibinfo{title}{How the geometry of cities
  determines urban scaling laws}}.
\newblock {\emph{\JournalTitle{Journal of the Royal Society Interface}}}
  \textbf{\bibinfo{volume}{18}}, \bibinfo{pages}{20200705}
  (\bibinfo{year}{2021}).

\bibitem{currid2010two}
\bibinfo{author}{Currid, E.} \& \bibinfo{author}{Williams, S.}
\newblock \bibinfo{journal}{\bibinfo{title}{Two cities, five industries:
  Similarities and differences within and between cultural industries in new
  york and los angeles}}.
\newblock {\emph{\JournalTitle{Journal of Planning Education and Research}}}
  \textbf{\bibinfo{volume}{29}}, \bibinfo{pages}{322--335}
  (\bibinfo{year}{2010}).

\bibitem{cheng2019network}
\bibinfo{author}{Cheng, F.}, \bibinfo{author}{Kov{\'a}cs, I.~A.} \&
  \bibinfo{author}{Barab{\'a}si, A.-L.}
\newblock \bibinfo{journal}{\bibinfo{title}{Network-based prediction of drug
  combinations}}.
\newblock {\emph{\JournalTitle{Nature communications}}}
  \textbf{\bibinfo{volume}{10}}, \bibinfo{pages}{1--11} (\bibinfo{year}{2019}).

\bibitem{jalili2017link}
\bibinfo{author}{Jalili, M.}, \bibinfo{author}{Orouskhani, Y.},
  \bibinfo{author}{Asgari, M.}, \bibinfo{author}{Alipourfard, N.} \&
  \bibinfo{author}{Perc, M.}
\newblock \bibinfo{journal}{\bibinfo{title}{Link prediction in multiplex online
  social networks}}.
\newblock {\emph{\JournalTitle{Royal Society open science}}}
  \textbf{\bibinfo{volume}{4}}, \bibinfo{pages}{160863} (\bibinfo{year}{2017}).

\bibitem{lerique2020joint}
\bibinfo{author}{Lerique, S.}, \bibinfo{author}{Abitbol, J.~L.} \&
  \bibinfo{author}{Karsai, M.}
\newblock \bibinfo{journal}{\bibinfo{title}{Joint embedding of structure and
  features via graph convolutional networks}}.
\newblock {\emph{\JournalTitle{Applied Network Science}}}
  \textbf{\bibinfo{volume}{5}}, \bibinfo{pages}{1--24} (\bibinfo{year}{2020}).

\bibitem{ren2014predicting}
\bibinfo{author}{Ren, Y.}, \bibinfo{author}{Ercsey-Ravasz, M.},
  \bibinfo{author}{Wang, P.}, \bibinfo{author}{Gonz{\'a}lez, M.~C.} \&
  \bibinfo{author}{Toroczkai, Z.}
\newblock \bibinfo{journal}{\bibinfo{title}{Predicting commuter flows in
  spatial networks using a radiation model based on temporal ranges}}.
\newblock {\emph{\JournalTitle{Nature communications}}}
  \textbf{\bibinfo{volume}{5}}, \bibinfo{pages}{1--9} (\bibinfo{year}{2014}).

\bibitem{teney2017graph}
\bibinfo{author}{Teney, D.}, \bibinfo{author}{Liu, L.} \& \bibinfo{author}{van
  Den~Hengel, A.}
\newblock \bibinfo{title}{Graph-structured representations for visual question
  answering}.
\newblock In \emph{\bibinfo{booktitle}{Proceedings of the IEEE conference on
  computer vision and pattern recognition}}, \bibinfo{pages}{1--9}
  (\bibinfo{year}{2017}).

\bibitem{wu2020learning}
\bibinfo{author}{Wu, N.}, \bibinfo{author}{Zhao, X.~W.}, \bibinfo{author}{Wang,
  J.} \& \bibinfo{author}{Pan, D.}
\newblock \bibinfo{title}{Learning effective road network representation with
  hierarchical graph neural networks}.
\newblock In \emph{\bibinfo{booktitle}{Proceedings of the 26th ACM SIGKDD
  International Conference on Knowledge Discovery \& Data Mining}},
  \bibinfo{pages}{6--14} (\bibinfo{year}{2020}).

\bibitem{gebru2017using}
\bibinfo{author}{Gebru, T.} \emph{et~al.}
\newblock \bibinfo{journal}{\bibinfo{title}{Using deep learning and google
  street view to estimate the demographic makeup of neighborhoods across the
  united states}}.
\newblock {\emph{\JournalTitle{Proceedings of the National Academy of
  Sciences}}} \textbf{\bibinfo{volume}{114}}, \bibinfo{pages}{13108--13113}
  (\bibinfo{year}{2017}).

\bibitem{abitbol2020interpretable}
\bibinfo{author}{Abitbol, J.~L.} \& \bibinfo{author}{Karsai, M.}
\newblock \bibinfo{journal}{\bibinfo{title}{Interpretable socioeconomic status
  inference from aerial imagery through urban patterns}}.
\newblock {\emph{\JournalTitle{Nature Machine Intelligence}}}
  \textbf{\bibinfo{volume}{2}}, \bibinfo{pages}{684--692}
  (\bibinfo{year}{2020}).

\bibitem{kempinska2019modelling}
\bibinfo{author}{Kempinska, K.} \& \bibinfo{author}{Murcio, R.}
\newblock \bibinfo{journal}{\bibinfo{title}{Modelling urban networks using
  variational autoencoders}}.
\newblock {\emph{\JournalTitle{Applied Network Science}}}
  \textbf{\bibinfo{volume}{4}}, \bibinfo{pages}{1--11} (\bibinfo{year}{2019}).

\bibitem{peng2011urbanization}
\bibinfo{author}{Peng, X.}, \bibinfo{author}{Chen, X.} \&
  \bibinfo{author}{Cheng, Y.}
\newblock \emph{\bibinfo{title}{Urbanization and its consequences}}
  (\bibinfo{publisher}{Paris, France: Eolss Publishers}, \bibinfo{year}{2011}).

\bibitem{hanson2004geography}
\bibinfo{author}{Hanson, S.} \& \bibinfo{author}{Giuliano, G.}
\newblock \emph{\bibinfo{title}{The geography of urban transportation}}
  (\bibinfo{publisher}{Guilford Press}, \bibinfo{year}{2004}).

\bibitem{cook2008mobilising}
\bibinfo{author}{Cook, I.~R.}
\newblock \bibinfo{journal}{\bibinfo{title}{Mobilising urban policies: The
  policy transfer of us business improvement districts to england and wales}}.
\newblock {\emph{\JournalTitle{Urban Studies}}} \textbf{\bibinfo{volume}{45}},
  \bibinfo{pages}{773--795} (\bibinfo{year}{2008}).

\bibitem{ghasemian2020stacking}
\bibinfo{author}{Ghasemian, A.}, \bibinfo{author}{Hosseinmardi, H.},
  \bibinfo{author}{Galstyan, A.}, \bibinfo{author}{Airoldi, E.~M.} \&
  \bibinfo{author}{Clauset, A.}
\newblock \bibinfo{journal}{\bibinfo{title}{Stacking models for nearly optimal
  link prediction in complex networks}}.
\newblock {\emph{\JournalTitle{Proceedings of the National Academy of
  Sciences}}} \textbf{\bibinfo{volume}{117}}, \bibinfo{pages}{23393--23400}
  (\bibinfo{year}{2020}).

\bibitem{clauset2008hierarchical}
\bibinfo{author}{Clauset, A.}, \bibinfo{author}{Moore, C.} \&
  \bibinfo{author}{Newman, M.~E.}
\newblock \bibinfo{journal}{\bibinfo{title}{Hierarchical structure and the
  prediction of missing links in networks}}.
\newblock {\emph{\JournalTitle{Nature}}} \textbf{\bibinfo{volume}{453}},
  \bibinfo{pages}{98--101} (\bibinfo{year}{2008}).

\bibitem{stanfield2017drug}
\bibinfo{author}{Stanfield, Z.}, \bibinfo{author}{Co{\c{s}}kun, M.} \&
  \bibinfo{author}{Koyut{\"u}rk, M.}
\newblock \bibinfo{journal}{\bibinfo{title}{Drug response prediction as a link
  prediction problem}}.
\newblock {\emph{\JournalTitle{Scientific reports}}}
  \textbf{\bibinfo{volume}{7}}, \bibinfo{pages}{1--13} (\bibinfo{year}{2017}).

\bibitem{schlichtkrull2018modeling}
\bibinfo{author}{Schlichtkrull, M.} \emph{et~al.}
\newblock \bibinfo{title}{Modeling relational data with graph convolutional
  networks}.
\newblock In \emph{\bibinfo{booktitle}{European semantic web conference}},
  \bibinfo{pages}{593--607} (\bibinfo{organization}{Springer},
  \bibinfo{year}{2018}).

\bibitem{barrington2019global}
\bibinfo{author}{Barrington-Leigh, C.} \& \bibinfo{author}{Millard-Ball, A.}
\newblock \bibinfo{journal}{\bibinfo{title}{A global assessment of
  street-network sprawl}}.
\newblock {\emph{\JournalTitle{PloS one}}} \textbf{\bibinfo{volume}{14}},
  \bibinfo{pages}{e0223078} (\bibinfo{year}{2019}).

\bibitem{barrington2020global}
\bibinfo{author}{Barrington-Leigh, C.} \& \bibinfo{author}{Millard-Ball, A.}
\newblock \bibinfo{journal}{\bibinfo{title}{Global trends toward urban
  street-network sprawl}}.
\newblock {\emph{\JournalTitle{Proceedings of the National Academy of
  Sciences}}} \textbf{\bibinfo{volume}{117}}, \bibinfo{pages}{1941--1950}
  (\bibinfo{year}{2020}).

\bibitem{hammack1984philadelphia}
\bibinfo{author}{Hammack, D.~C.}
\newblock \bibinfo{title}{Philadelphia: A 300-year history}
  (\bibinfo{year}{1984}).

\bibitem{barthelemy2013self}
\bibinfo{author}{Barthelemy, M.}, \bibinfo{author}{Bordin, P.},
  \bibinfo{author}{Berestycki, H.} \& \bibinfo{author}{Gribaudi, M.}
\newblock \bibinfo{journal}{\bibinfo{title}{Self-organization versus top-down
  planning in the evolution of a city}}.
\newblock {\emph{\JournalTitle{Scientific reports}}}
  \textbf{\bibinfo{volume}{3}}, \bibinfo{pages}{1--8} (\bibinfo{year}{2013}).

\bibitem{peterson2009birth}
\bibinfo{author}{Peterson, J.~A.}
\newblock \bibinfo{journal}{\bibinfo{title}{The birth of organized city
  planning in the united states, 1909--1910}}.
\newblock {\emph{\JournalTitle{Journal of the American Planning Association}}}
  \textbf{\bibinfo{volume}{75}}, \bibinfo{pages}{123--133}
  (\bibinfo{year}{2009}).

\bibitem{boeing2020multi}
\bibinfo{author}{Boeing, G.}
\newblock \bibinfo{journal}{\bibinfo{title}{A multi-scale analysis of 27,000
  urban street networks: Every us city, town, urbanized area, and zillow
  neighborhood}}.
\newblock {\emph{\JournalTitle{Environment and Planning B: Urban Analytics and
  City Science}}} \textbf{\bibinfo{volume}{47}}, \bibinfo{pages}{590--608}
  (\bibinfo{year}{2020}).

\bibitem{wang2015resilience}
\bibinfo{author}{Wang, J.}
\newblock \bibinfo{journal}{\bibinfo{title}{Resilience of self-organised and
  top-down planned cities—a case study on london and beijing street
  networks}}.
\newblock {\emph{\JournalTitle{PloS one}}} \textbf{\bibinfo{volume}{10}},
  \bibinfo{pages}{e0141736} (\bibinfo{year}{2015}).

\bibitem{giacomin2015road}
\bibinfo{author}{Giacomin, D.~J.} \& \bibinfo{author}{Levinson, D.~M.}
\newblock \bibinfo{journal}{\bibinfo{title}{Road network circuity in
  metropolitan areas}}.
\newblock {\emph{\JournalTitle{Environment and Planning B: Planning and
  Design}}} \textbf{\bibinfo{volume}{42}}, \bibinfo{pages}{1040--1053}
  (\bibinfo{year}{2015}).

\bibitem{ortman2014pre}
\bibinfo{author}{Ortman, S.~G.}, \bibinfo{author}{Cabaniss, A.~H.},
  \bibinfo{author}{Sturm, J.~O.} \& \bibinfo{author}{Bettencourt, L.~M.}
\newblock \bibinfo{journal}{\bibinfo{title}{The pre-history of urban scaling}}.
\newblock {\emph{\JournalTitle{PloS one}}} \textbf{\bibinfo{volume}{9}},
  \bibinfo{pages}{e87902} (\bibinfo{year}{2014}).

\bibitem{whittemore2012zoning}
\bibinfo{author}{Whittemore, A.~H.}
\newblock \bibinfo{journal}{\bibinfo{title}{Zoning los angeles: a brief history
  of four regimes}}.
\newblock {\emph{\JournalTitle{Planning Perspectives}}}
  \textbf{\bibinfo{volume}{27}}, \bibinfo{pages}{393--415}
  (\bibinfo{year}{2012}).

\bibitem{endoh2004historical}
\bibinfo{author}{ENDOH, T.}
\newblock \bibinfo{journal}{\bibinfo{title}{Historical review of reclamation
  works in tokyo port area}}.
\newblock {\emph{\JournalTitle{Journal of Geography (Chigaku Zasshi)}}}
  \textbf{\bibinfo{volume}{113}}, \bibinfo{pages}{534--538}
  (\bibinfo{year}{2004}).

\bibitem{bettencourt2020urban}
\bibinfo{author}{Bettencourt, L.~M.}
\newblock \bibinfo{journal}{\bibinfo{title}{Urban growth and the emergent
  statistics of cities}}.
\newblock {\emph{\JournalTitle{Science advances}}}
  \textbf{\bibinfo{volume}{6}}, \bibinfo{pages}{eaat8812}
  (\bibinfo{year}{2020}).

\bibitem{wei2016transfer}
\bibinfo{author}{Wei, Y.}, \bibinfo{author}{Zheng, Y.} \&
  \bibinfo{author}{Yang, Q.}
\newblock \bibinfo{title}{Transfer knowledge between cities}.
\newblock In \emph{\bibinfo{booktitle}{Proceedings of the 22nd ACM SIGKDD
  International Conference on Knowledge Discovery and Data Mining}},
  \bibinfo{pages}{1905--1914} (\bibinfo{year}{2016}).

\bibitem{dai2009eigentransfer}
\bibinfo{author}{Dai, W.}, \bibinfo{author}{Jin, O.}, \bibinfo{author}{Xue,
  G.-R.}, \bibinfo{author}{Yang, Q.} \& \bibinfo{author}{Yu, Y.}
\newblock \bibinfo{title}{Eigentransfer: a unified framework for transfer
  learning}.
\newblock In \emph{\bibinfo{booktitle}{Proceedings of the 26th Annual
  International Conference on Machine Learning}}, \bibinfo{pages}{193--200}
  (\bibinfo{year}{2009}).

\bibitem{dong2019predicting}
\bibinfo{author}{Dong, L.}, \bibinfo{author}{Ratti, C.} \&
  \bibinfo{author}{Zheng, S.}
\newblock \bibinfo{journal}{\bibinfo{title}{Predicting neighborhoods’
  socioeconomic attributes using restaurant data}}.
\newblock {\emph{\JournalTitle{Proceedings of the National Academy of
  Sciences}}} \textbf{\bibinfo{volume}{116}}, \bibinfo{pages}{15447--15452}
  (\bibinfo{year}{2019}).

\bibitem{mandelbrot1982fractal}
\bibinfo{author}{Mandelbrot, B.~B.}
\newblock \emph{\bibinfo{title}{The fractal geometry of nature}},
  vol.~\bibinfo{volume}{1} (\bibinfo{year}{1982}).

\bibitem{falconer1999techniques}
\bibinfo{author}{Falconer, K.}
\newblock \bibinfo{journal}{\bibinfo{title}{Techniques in fractal geometry}}.
\newblock {\emph{\JournalTitle{ADDITIVE NUMBER THEORY: THE CLASSICAL BASES}}}
  \textbf{\bibinfo{volume}{31}}, \bibinfo{pages}{119} (\bibinfo{year}{1999}).

\bibitem{meakin1983formation}
\bibinfo{author}{Meakin, P.}
\newblock \bibinfo{journal}{\bibinfo{title}{Formation of fractal clusters and
  networks by irreversible diffusion-limited aggregation}}.
\newblock {\emph{\JournalTitle{Physical Review Letters}}}
  \textbf{\bibinfo{volume}{51}}, \bibinfo{pages}{1119} (\bibinfo{year}{1983}).

\bibitem{batty1994fractal}
\bibinfo{author}{Batty, M.} \& \bibinfo{author}{Longley, P.~A.}
\newblock \emph{\bibinfo{title}{Fractal cities: a geometry of form and
  function}} (\bibinfo{publisher}{Academic press}, \bibinfo{year}{1994}).

\bibitem{sidqi2019comparing}
\bibinfo{author}{Sidqi, Y.}, \bibinfo{author}{Thomas, I.},
  \bibinfo{author}{Frankhauser, P.} \& \bibinfo{author}{Reti{\`e}re, N.}
\newblock \bibinfo{journal}{\bibinfo{title}{Comparing fractal indices of
  electric networks to roads and buildings: The case of grenoble (france)}}.
\newblock {\emph{\JournalTitle{Physica A: Statistical Mechanics and its
  Applications}}} \textbf{\bibinfo{volume}{531}}, \bibinfo{pages}{121774}
  (\bibinfo{year}{2019}).

\bibitem{ariza2013multifractal}
\bibinfo{author}{Ariza-Villaverde, A.~B.},
  \bibinfo{author}{Jim{\'e}nez-Hornero, F.~J.} \& \bibinfo{author}{De~Rav{\'e},
  E.~G.}
\newblock \bibinfo{journal}{\bibinfo{title}{Multifractal analysis of axial maps
  applied to the study of urban morphology}}.
\newblock {\emph{\JournalTitle{Computers, Environment and Urban Systems}}}
  \textbf{\bibinfo{volume}{38}}, \bibinfo{pages}{1--10} (\bibinfo{year}{2013}).

\bibitem{makse1998modeling}
\bibinfo{author}{Makse, H.~A.} \emph{et~al.}
\newblock \bibinfo{journal}{\bibinfo{title}{Modeling urban growth patterns with
  correlated percolation}}.
\newblock {\emph{\JournalTitle{Physical Review E}}}
  \textbf{\bibinfo{volume}{58}}, \bibinfo{pages}{7054} (\bibinfo{year}{1998}).

\bibitem{murcio2015multifractal}
\bibinfo{author}{Murcio, R.}, \bibinfo{author}{Masucci, A.~P.},
  \bibinfo{author}{Arcaute, E.} \& \bibinfo{author}{Batty, M.}
\newblock \bibinfo{journal}{\bibinfo{title}{Multifractal to monofractal
  evolution of the london street network}}.
\newblock {\emph{\JournalTitle{Physical Review E}}}
  \textbf{\bibinfo{volume}{92}}, \bibinfo{pages}{062130}
  (\bibinfo{year}{2015}).

\bibitem{he2016deep}
\bibinfo{author}{He, K.}, \bibinfo{author}{Zhang, X.}, \bibinfo{author}{Ren,
  S.} \& \bibinfo{author}{Sun, J.}
\newblock \bibinfo{title}{Deep residual learning for image recognition}.
\newblock In \emph{\bibinfo{booktitle}{Proceedings of the IEEE conference on
  computer vision and pattern recognition}}, \bibinfo{pages}{770--778}
  (\bibinfo{year}{2016}).

\bibitem{chen2020measuring}
\bibinfo{author}{Chen, D.} \emph{et~al.}
\newblock \bibinfo{title}{Measuring and relieving the over-smoothing problem
  for graph neural networks from the topological view}.
\newblock In \emph{\bibinfo{booktitle}{Proceedings of the AAAI Conference on
  Artificial Intelligence}}, vol.~\bibinfo{volume}{34},
  \bibinfo{pages}{3438--3445} (\bibinfo{year}{2020}).

\bibitem{berry1969metropolitan}
\bibinfo{author}{Berry, B.~J.}, \bibinfo{author}{Goheen, P.~G.} \&
  \bibinfo{author}{Goldstein, H.}
\newblock \emph{\bibinfo{title}{Metropolitan area definition: A re-evaluation
  of concept and statistical practice}}, vol.~\bibinfo{volume}{28}
  (\bibinfo{publisher}{[Washington]: US Bureau of the Census},
  \bibinfo{year}{1969}).

\bibitem{corbane2019automated}
\bibinfo{author}{Corbane, C.} \emph{et~al.}
\newblock \bibinfo{journal}{\bibinfo{title}{Automated global delineation of
  human settlements from 40 years of landsat satellite data archives}}.
\newblock {\emph{\JournalTitle{Big Earth Data}}} \textbf{\bibinfo{volume}{3}},
  \bibinfo{pages}{140--169} (\bibinfo{year}{2019}).

\bibitem{rozenfeld2008laws}
\bibinfo{author}{Rozenfeld, H.~D.} \emph{et~al.}
\newblock \bibinfo{journal}{\bibinfo{title}{Laws of population growth}}.
\newblock {\emph{\JournalTitle{Proceedings of the National Academy of
  Sciences}}} \textbf{\bibinfo{volume}{105}}, \bibinfo{pages}{18702--18707}
  (\bibinfo{year}{2008}).

\bibitem{shen2019delineating}
\bibinfo{author}{Shen, Y.} \& \bibinfo{author}{Batty, M.}
\newblock \bibinfo{journal}{\bibinfo{title}{Delineating the perceived
  functional regions of london from commuting flows}}.
\newblock {\emph{\JournalTitle{Environment and Planning A: Economy and Space}}}
  \textbf{\bibinfo{volume}{51}}, \bibinfo{pages}{547--550}
  (\bibinfo{year}{2019}).

\bibitem{long2016mapping}
\bibinfo{author}{Long, Y.}, \bibinfo{author}{Shen, Y.} \& \bibinfo{author}{Jin,
  X.}
\newblock \bibinfo{journal}{\bibinfo{title}{Mapping block-level urban areas for
  all chinese cities}}.
\newblock {\emph{\JournalTitle{Annals of the American Association of
  Geographers}}} \textbf{\bibinfo{volume}{106}}, \bibinfo{pages}{96--113}
  (\bibinfo{year}{2016}).

\bibitem{cao2020quantifying}
\bibinfo{author}{Cao, W.}, \bibinfo{author}{Dong, L.}, \bibinfo{author}{Wu, L.}
  \& \bibinfo{author}{Liu, Y.}
\newblock \bibinfo{journal}{\bibinfo{title}{Quantifying urban areas with
  multi-source data based on percolation theory}}.
\newblock {\emph{\JournalTitle{Remote Sensing of Environment}}}
  \textbf{\bibinfo{volume}{241}}, \bibinfo{pages}{111730}
  (\bibinfo{year}{2020}).

\bibitem{zischg2019century}
\bibinfo{author}{Zischg, J.}, \bibinfo{author}{Klinkhamer, C.},
  \bibinfo{author}{Zhan, X.}, \bibinfo{author}{Rao, P. S.~C.} \&
  \bibinfo{author}{Sitzenfrei, R.}
\newblock \bibinfo{journal}{\bibinfo{title}{A century of topological
  coevolution of complex infrastructure networks in an alpine city}}.
\newblock {\emph{\JournalTitle{Complexity}}}  (\bibinfo{year}{2019}).

\bibitem{yabe2020unsupervised}
\bibinfo{author}{Yabe, T.}, \bibinfo{author}{Tsubouchi, K.},
  \bibinfo{author}{Shimizu, T.}, \bibinfo{author}{Sekimoto, Y.} \&
  \bibinfo{author}{Ukkusuri, S.~V.}
\newblock \bibinfo{title}{Unsupervised translation via hierarchical anchoring:
  Functional mapping of places across cities}.
\newblock In \emph{\bibinfo{booktitle}{Proceedings of the 26th ACM SIGKDD
  International Conference on Knowledge Discovery \& Data Mining}},
  \bibinfo{pages}{2841--2851} (\bibinfo{year}{2020}).

\bibitem{zhao2016spatio}
\bibinfo{author}{Zhao, J.}, \bibinfo{author}{Li, D.},
  \bibinfo{author}{Sanhedrai, H.}, \bibinfo{author}{Cohen, R.} \&
  \bibinfo{author}{Havlin, S.}
\newblock \bibinfo{journal}{\bibinfo{title}{Spatio-temporal propagation of
  cascading overload failures in spatially embedded networks}}.
\newblock {\emph{\JournalTitle{Nature communications}}}
  \textbf{\bibinfo{volume}{7}}, \bibinfo{pages}{1--6} (\bibinfo{year}{2016}).

\bibitem{loder2019understanding}
\bibinfo{author}{Loder, A.}, \bibinfo{author}{Amb{\"u}hl, L.},
  \bibinfo{author}{Menendez, M.} \& \bibinfo{author}{Axhausen, K.~W.}
\newblock \bibinfo{journal}{\bibinfo{title}{Understanding traffic capacity of
  urban networks}}.
\newblock {\emph{\JournalTitle{Scientific reports}}}
  \textbf{\bibinfo{volume}{9}}, \bibinfo{pages}{1--10} (\bibinfo{year}{2019}).

\bibitem{zeng2020multiple}
\bibinfo{author}{Zeng, G.} \emph{et~al.}
\newblock \bibinfo{journal}{\bibinfo{title}{Multiple metastable network states
  in urban traffic}}.
\newblock {\emph{\JournalTitle{Proceedings of the National Academy of
  Sciences}}} \textbf{\bibinfo{volume}{117}}, \bibinfo{pages}{17528--17534}
  (\bibinfo{year}{2020}).

\bibitem{boeing2017osmnx}
\bibinfo{author}{Boeing, G.}
\newblock \bibinfo{journal}{\bibinfo{title}{Osmnx: New methods for acquiring,
  constructing, analyzing, and visualizing complex street networks}}.
\newblock {\emph{\JournalTitle{Computers, Environment and Urban Systems}}}
  \textbf{\bibinfo{volume}{65}}, \bibinfo{pages}{126--139}
  (\bibinfo{year}{2017}).

\bibitem{ganin2017resilience}
\bibinfo{author}{Ganin, A.~A.} \emph{et~al.}
\newblock \bibinfo{journal}{\bibinfo{title}{Resilience and efficiency in
  transportation networks}}.
\newblock {\emph{\JournalTitle{Science advances}}}
  \textbf{\bibinfo{volume}{3}}, \bibinfo{pages}{e1701079}
  (\bibinfo{year}{2017}).

\bibitem{louail2014mobile}
\bibinfo{author}{Louail, T.} \emph{et~al.}
\newblock \bibinfo{journal}{\bibinfo{title}{From mobile phone data to the
  spatial structure of cities}}.
\newblock {\emph{\JournalTitle{Scientific reports}}}
  \textbf{\bibinfo{volume}{4}}, \bibinfo{pages}{1--12} (\bibinfo{year}{2014}).

\bibitem{thompson2020global}
\bibinfo{author}{Thompson, J.} \emph{et~al.}
\newblock \bibinfo{journal}{\bibinfo{title}{A global analysis of urban design
  types and road transport injury: an image processing study}}.
\newblock {\emph{\JournalTitle{The Lancet Planetary Health}}}
  \textbf{\bibinfo{volume}{4}}, \bibinfo{pages}{e32--e42}
  (\bibinfo{year}{2020}).

\bibitem{ua2012}
\bibinfo{author}{Agency, E.~E.}
\newblock \bibinfo{title}{Urban atlas 2012} (\bibinfo{year}{accessed July
  2021}).
\newblock \bibinfo{note}{Available online at
  \url{https://land.copernicus.eu/local/urban-atlas/urban-atlas-2012}}.

\bibitem{khiali2019combining}
\bibinfo{author}{Khiali-Miab, A.}, \bibinfo{author}{van Strien, M.~J.},
  \bibinfo{author}{Axhausen, K.~W.} \& \bibinfo{author}{Gr{\^e}t-Regamey, A.}
\newblock \bibinfo{journal}{\bibinfo{title}{Combining urban scaling and
  polycentricity to explain socio-economic status of urban regions}}.
\newblock {\emph{\JournalTitle{PloS one}}} \textbf{\bibinfo{volume}{14}},
  \bibinfo{pages}{e0218022} (\bibinfo{year}{2019}).

\bibitem{rozenblat2020extending}
\bibinfo{author}{Rozenblat, C.}
\newblock \bibinfo{journal}{\bibinfo{title}{Extending the concept of city for
  delineating large urban regions (lur) for the cities of the world}}.
\newblock {\emph{\JournalTitle{Cybergeo: European Journal of Geography}}}
  (\bibinfo{year}{2020}).

\bibitem{ma2020functional}
\bibinfo{author}{Ma, S.} \& \bibinfo{author}{Long, Y.}
\newblock \bibinfo{journal}{\bibinfo{title}{Functional urban area delineations
  of cities on the chinese mainland using massive didi ride-hailing records}}.
\newblock {\emph{\JournalTitle{Cities}}} \textbf{\bibinfo{volume}{97}},
  \bibinfo{pages}{102532} (\bibinfo{year}{2020}).

\bibitem{yang2014embedding}
\bibinfo{author}{Yang, B.}, \bibinfo{author}{Yih, W.-t.}, \bibinfo{author}{He,
  X.}, \bibinfo{author}{Gao, J.} \& \bibinfo{author}{Deng, L.}
\newblock \bibinfo{journal}{\bibinfo{title}{Embedding entities and relations
  for learning and inference in knowledge bases}}.
\newblock {\emph{\JournalTitle{arXiv preprint arXiv:1412.6575}}}
  (\bibinfo{year}{2014}).

\bibitem{grover2016node2vec}
\bibinfo{author}{Grover, A.} \& \bibinfo{author}{Leskovec, J.}
\newblock \bibinfo{title}{node2vec: Scalable feature learning for networks}.
\newblock In \emph{\bibinfo{booktitle}{Proceedings of the 22nd ACM SIGKDD
  international conference on Knowledge discovery and data mining}},
  \bibinfo{pages}{855--864} (\bibinfo{year}{2016}).

\bibitem{ribeiro2017struc2vec}
\bibinfo{author}{Ribeiro, L.~F.}, \bibinfo{author}{Saverese, P.~H.} \&
  \bibinfo{author}{Figueiredo, D.~R.}
\newblock \bibinfo{title}{struc2vec: Learning node representations from
  structural identity}.
\newblock In \emph{\bibinfo{booktitle}{Proceedings of the 23rd ACM SIGKDD
  international conference on knowledge discovery and data mining}},
  \bibinfo{pages}{385--394} (\bibinfo{year}{2017}).

\bibitem{kipf2016semi}
\bibinfo{author}{Kipf, T.~N.} \& \bibinfo{author}{Welling, M.}
\newblock \bibinfo{journal}{\bibinfo{title}{Semi-supervised classification with
  graph convolutional networks}}.
\newblock {\emph{\JournalTitle{arXiv preprint arXiv:1609.02907}}}
  (\bibinfo{year}{2016}).

\bibitem{hamilton2017inductive}
\bibinfo{author}{Hamilton, W.~L.}, \bibinfo{author}{Ying, R.} \&
  \bibinfo{author}{Leskovec, J.}
\newblock \bibinfo{journal}{\bibinfo{title}{Inductive representation learning
  on large graphs}}.
\newblock {\emph{\JournalTitle{arXiv preprint arXiv:1706.02216}}}
  (\bibinfo{year}{2017}).

\bibitem{velivckovic2017graph}
\bibinfo{author}{Veli{\v{c}}kovi{\'c}, P.} \emph{et~al.}
\newblock \bibinfo{journal}{\bibinfo{title}{Graph attention networks}}.
\newblock {\emph{\JournalTitle{arXiv preprint arXiv:1710.10903}}}
  (\bibinfo{year}{2017}).

\bibitem{kingma2014adam}
\bibinfo{author}{Kingma, D.~P.} \& \bibinfo{author}{Ba, J.}
\newblock \bibinfo{journal}{\bibinfo{title}{Adam: A method for stochastic
  optimization}}.
\newblock {\emph{\JournalTitle{arXiv preprint arXiv:1412.6980}}}
  (\bibinfo{year}{2014}).

\end{thebibliography}
\end{document}